\documentclass{article}%
\usepackage{amsfonts}
\usepackage{amsmath}
\usepackage{amssymb}
\usepackage{graphicx}%
\newtheorem{theorem}{Theorem}

\newtheorem{corollary}[theorem]{Corollary}

\newtheorem{definition}[theorem]{Definition}

\newtheorem{lemma}[theorem]{Lemma}

\newtheorem{proposition}[theorem]{Proposition}
\newtheorem{remark}[theorem]{Remark}

\newenvironment{proof}[1][Proof]{\noindent\textbf{#1.} }{\ \rule{0.5em}{0.5em}}
\newcommand{\bpartial}{\mathop{\partial\kern -4pt\raisebox{.8pt}{$|$}}}
\newcommand{\bra}{\mathopen{[\kern-1.6pt[}}
\newcommand{\ket}{\mathclose{]\kern-1.5pt]}}
\newcommand{\bbra}{\mathopen{[\kern-2.2pt[\kern-2.3pt[}}
\newcommand{\bket}{\mathclose{]\kern-2.1pt]\kern-2.3pt]}}

\makeindex
\begin{document}

\title{Conservation Laws on Riemann-Cartan, Lorentzian and Teleparallel
Spacetimes\thanks{The contents of this paper appeared in two parts in
\textit{Bull. Soc. Sci. Lodz. Series }\textbf{57, }Ser.\textit{ Res. on
Deformations}, \textbf{52}, 37-65, 66-77 (2007). In the present version some
misprints of the published version have been corrected.}}
\author{\hspace{-1cm}{\small W. A. Rodrigues Jr.} {\small and Q. A. G. de Souza}\\$\hspace{-0.1cm}${\footnotesize Institute of Mathematics, Statistics and
Scientific Computation}\\{\footnotesize IMECC-UNICAMP CP 6065}\\{\footnotesize 13083-859 Campinas, SP, Brazil}\\{\small and}
\and {\small R. da Rocha}\\{\footnotesize Instituto de F\'{\i}sica Te\'{o}rica, UNESP, Rua Pamplona 145}\\{\footnotesize 01405-900, S\~{a}o Paulo, SP, Brazil.}\\{\footnotesize and}\\{\footnotesize DRCC - Institute of Physics Gleb Wataghin, UNICAMP CP 6165}\\{\footnotesize 13083-970 Campinas, SP, Brazil}\\{\footnotesize e-mails: \texttt{walrod@ime.unicamp.br; quin@ime.unicamp.br;
roldao@ifi.unicamp.br}}}
\maketitle

\begin{abstract}
Using a Clifford bundle formalism, we examine: (a)~the strong conditions for
existence of conservation laws involving only the energy-momentum and angular
momentum of the matter fields on a general Riemann-Cartan spacetime and the
particular cases of Lorentzian and teleparallel spacetimes and (b)~the
conditions for the existence of conservation laws of energy-momentum and
angular momentum for the matter and gravitational fields when this later
concept can be rigorously defined. We examine in more details some statements
concerning the issues of the conservation laws in General Relativity and
Riemann-Cartan (including the particular case of the teleparallel ones) theories.

\end{abstract}
\tableofcontents

\section{Introduction}

Using the Clifford bundle formalism of differential forms (see Appendix
A\footnote{In Appendix A we give a very short introduction to the main tools
of the the Clifford bundle formalism needed for this paper. A detailed and up
to date presentation to the Clifford bundle formalism is given, e.g., in
\cite{rodoliv2006}.}) we reexamine the origin and meaning of conservation laws
of energy-momentum and angular momentum and the conditions for their existence
on a general Riemann-Cartan spacetime (RCST)\footnote{See details in Appendix
A.{}} $(M,\mathbf{g},\nabla,\tau_{\mathbf{g}},\uparrow)$ and also in the
particular cases of Lorentzian spacetimes $\mathfrak{M}=(M,\mathbf{g}%
,D,\tau_{\mathbf{g}},\uparrow)$ which as it is well known model gravitational
fields in the General Relativity Theory (GRT) \cite{sawu}. A RCST is supposed
to model a \textit{generalized} gravitational field in the so called
Riemann-Cartan theories \cite{hehl}. The case of the so called
teleparallel\footnote{A teleparallel spacetime is a particular Riemann-Cartan
spacetime with null curvature and non null torsion tensor \cite{pe1,pe2,pe3}.}
equivalent of GRT \cite{maluf} is also investigated and the recent claim
\cite{deandrade} that there is a genuine energy-momentum conservation law in
that theory is investigated in more details.

In what follows, we suppose that a set of dynamic fields live and interact in
$(M,\mathbf{g},\nabla,\tau_{\mathbf{g}},\uparrow)$ (or $\mathfrak{M}$). Of
course, we want that the RCST admits spinor fields, which implies according to
Geroch's theorem that the orthonormal frame bundle must be\textit{ trivial
}\cite{geroch,moro,rodoliv2006}. This permits a great simplification in our
calculations, in particular if use is made of the calculation procedures of
the Clifford bundle formalism. Moreover, we will suppose, for simplicity that
the dynamic fields of the theory $\phi^{A}$, $A=1,2,...,n$, are $r$%
-forms\footnote{This is not a serious restriction in the formalism since as it
is shown in details in \cite{moro,rodoliv2006}, one can represent spinor
fields by sums of even multiform fields once a spinorial frame is given. The
functional derivative of non-homogeneous multiform fields is developed in
details in, e.g.,~\cite{rodoliv2006}.}, i.e., each $\phi^{A}\in\sec
\bigwedge^{r}T^{\ast}M\hookrightarrow\sec\mathcal{C}\ell(M,\mathtt{g})$, for
some $r=0,1,...,4$.

We recall the very important fact that there are in such theories a set of
`covariant conservation laws' which are identities which result from the fact
that Lagrangian densities of relativistic field theories are supposed to be
invariant under \textit{diffeomorphisms} and \textit{active} \textit{local
Lorentz rotations\footnote{Satisfying such a condition implies in general in
the use of generalized gauge connections, implying a sort of equivalence
between spacetimes equipped with connections having different curvature and/or
torsion tensors \cite{rodroldao,roldaowa}.}}. These covariant conservation
laws do \textit{not} express in general any genuine conservation law of
energy-momentum or angular momentum. We prove moreover, as first shown by
\cite{benn} that genuine conservation laws of energy-momentum and angular
momentum for only the \textit{matter} \textit{fields} exist for a field theory
in a RCST only if there exists a set of\footnote{The maximum possible value of
the integer number $m$ in a $4$-dimensional spacetime is ten.} $m$ appropriate
vector fields $\xi^{(a)}$, $a=1,2,...,m$ such that $\pounds _{\xi^{(a)}%
}\mathbf{g}=0$ and $\pounds _{\xi^{(a)}}\Theta=0$, where $\Theta$ is the
torsion tensor.

Thus, we show in Section 6 that in the teleparallel version of GRT, the
existence of Killing vector fields \textit{does not} warrant (contrary to the
case of GRT) the existence of conservation laws involving \textit{only} the
energy-momentum tensors of the \textit{matter} fields. We show moreover, still
in Section 6, that in the teleparallel version of GRT (with null or non null
cosmological constant) there is a genuine conservation law involving the
energy-momentum tensor of matter and the energy-momentum tensor of the
gravitational field, which in that theory is a well defined object.

Although this is a well known result, we think that our formalism puts it in a
new perspective. Indeed, in our approach, the teleparallel equivalent of
General Relativity as formulated, e.g., by \cite{maluf} or \cite{deandrade},
is easily seem as consisting in the introduction of: (a)~a bilinear form (a
deformed metric tensor \cite{roqui,rodoliv2006}) $\mathbf{g}=\eta
_{\mathbf{ab}}\theta^{\mathbf{a}}\otimes\theta^{\mathbf{b}}$, (b) a
teleparallel connection (necessary to make the theory invariant under active
local Lorentz transformations\footnote{On the issue on active local Lorentz
invariance, see also~\cite{roldaowa,rodroldao}.{}}) in the manifold~$M\simeq
\mathbb{R}^{4}$ of Minkowski spacetime structure, and (c)~a Lagrangian density
differing from the Einstein-Hilbert Lagrangian density by an exact differential.

The paper is organized as follows:

Section 2 and Appendix~A are aimed to give to the reader some background
information needed to better understand our developments. In Section~2 we
recall some mathematical preliminaries as the definition of vertical and
horizontal variations, the concept of functional derivatives of functionals on
a $1$-jet bundle, the Euler-Lagrange equations (ELE) and the fact that the
action of any theory formulated in terms of differential forms is invariant
under diffeomorphisms, whereas in Appendix~A we briefly describe the Clifford
bundle formalism used throughout the paper. Appendix A also provides a
derivation of the energy-momentum $3$-forms for the electromagnetic field
which in the Clifford bundle formalism (and our conventions) are expressed
very elegantly by $-\star T^{\mathbf{a}}=\star\mathcal{T}^{\mathbf{a}}%
=-\frac{1}{2}\star(F\theta^{\mathbf{a}}\tilde{F})$.

In Section~3 we recall the proof of a set of identities called `covariant
conservation laws' valid in a RCST \cite{benn}, which as already mentioned
above do not encode, in general, any genuine energy-momentum and/or angular
momentum conservation laws.

In Section~4 we assume that the Lagrangian density is invariant under active
local Lorentz transformations and diffeomorphisms and then recall the
conditions for the existence of genuine conservation laws in a RCST which
involve only the energy-momentum and angular momentum tensor of the
\textit{matter} fields \cite{benn}.

Next, in Section~5, we recall (for completeness) with our formalism the theory
of pseudo-potentials and pseudo energy-momentum tensors in GRT, and show that
there are in general no conservation laws of energy-momentum and angular
momentum in this theory \cite{rodoliv04}. We also discuss some misleading and
even wrong statements concerning this issue that appear in the literature.

Finally, in Section~6 we discuss the conservation laws in the teleparallel
equivalent of General Relativity, as already mentioned above.

Our conclusions can be found in Section~7. To better illustrate the meaning of
our results, we also present, in Appendix~B, various examples showing that not
all Killing vector fields of a teleparallel spacetime (Schwarzschild,
de~Sitter, Friedmann) satisfy Eq.(\ref{7.26}) meaning that in a model of the
teleparallel `equivalent' of GRT there are, in general, fewer conservation
laws involving only the matter fields than in the corresponding model of GRT.

\section{ Some Preliminaries}

\subsection{Variations}

\subsubsection{Vertical Variation}

Let $X\in\sec\mathcal{C}\ell(M,\mathtt{g})$, be a Clifford (multiform)
field\footnote{If $X=\psi\in\sec\mathcal{C}\ell^{(0)}(M,\mathtt{g})$ (where
$\mathcal{C}\ell^{(0)}(M,\mathtt{g})$ is the even subbundle of $\mathcal{C}%
\ell(M,\mathtt{g})$) is a \textit{representative} of a Dirac-Hestenes spinor
field in a given spin frame, then an active local transformation sends
$\psi\mapsto\psi^{\prime}$, with $\psi^{\prime}=L\psi$ \cite{rodoliv2006}.}.
An active local Lorentz transformation sends $X\mapsto X^{\prime}\in
\sec\mathcal{C}\ell(M,\mathtt{g})$, with
\begin{equation}
X^{\prime}=UX\tilde{U}. \label{6.13}%
\end{equation}

Each $U\in\sec\mathrm{Spin}_{1,3}^{e}(M)$ can be written (see, e.g.,
\cite{rodoliv2006}) as $\pm$ the exponential of a 2-form field $F\in
\sec\bigwedge\nolimits^{2}T^{\ast}M$ $\hookrightarrow\sec$ \ $\mathcal{C}%
\ell(M,\mathtt{g})$. For infinitesimal transformations we must choose the $+$
sign and write $F=\alpha f$, $\alpha\ll1$, $F^{2}\neq0$.

\begin{definition}
\label{var vertical}Let $X$ be a Clifford field. The vertical variation of $X$
is the field $%
\mbox{\boldmath{$\delta$}}%
_{v}X$ \emph{(}of the same nature of $X$\emph{)} such that%
\begin{equation}%
\mbox{\boldmath{$\delta$}}%
_{v}X=X^{\prime}-X. \label{6.12d}%
\end{equation}

\end{definition}

\begin{remark}
The case where $F$ is independent of $x$ $\in M$ is said to be a gauge
transformation of the first kind, and the general case is said to be a gauge
transformation of the second kind.
\end{remark}

\subsubsection{Horizontal Variation}

Let $\sigma_{t}$ be a one-parameter group of diffeomorphisms of $M$ and let
$\xi\in\sec TM$ be the vector field that generates $\sigma_{t}$, i.e.,%
\begin{equation}
\xi^{\mu}(x)=\left.  \frac{d\sigma_{t}^{\mu}(x)}{dt}\right\vert _{t=0}.
\label{6.15}%
\end{equation}

\begin{definition}
\label{var horizontal}We call the horizontal variation of $X$ induced by a
one-parameter group of diffeomorphisms of $M$ to be the quantity%
\begin{equation}%
\mbox{\boldmath{$\delta$}}%
_{h}X=\lim_{t\rightarrow0}\frac{\sigma_{t}^{\ast}X-X}{t}=-\pounds _{\xi}X.
\label{6.16}%
\end{equation}

\end{definition}

\begin{definition}
\label{var total}We call total variation of a multiform field $X$ to the
quantity%
\begin{equation}%
\mbox{\boldmath{$\delta$}}%
X=%
\mbox{\boldmath{$\delta$}}%
_{v}X+%
\mbox{\boldmath{$\delta$}}%
_{h}X=%
\mbox{\boldmath{$\delta$}}%
_{v}X-\pounds _{\xi}X. \label{6.17}%
\end{equation}

\end{definition}

It is \textit{crucial} to distinguish between the two variations defined above.

\subsection{Functional Derivatives}

Let $J^{1}(\bigwedge T^{\ast}M)$ be the 1-jet bundle over $\bigwedge T^{\ast
}M\hookrightarrow\mathcal{C}\ell(M,\mathtt{g})$, i.e., the vector bundle
defined by%
\begin{equation}
J^{1}(\bigwedge T^{\ast}M)=\{(x,\phi(x),d\phi(x))\text{; }x\in M\text{, }%
\phi\in\sec\bigwedge T^{\ast}M\hookrightarrow\sec\mathcal{C}\ell
(M,\mathtt{g})\}.
\end{equation}

Then, with each local section $\phi\in\sec\bigwedge T^{\ast}M\hookrightarrow
\sec\mathcal{C}\ell(M,\mathtt{g})\}$, we may associate a local section
$j_{1}(\phi)\in\sec J^{1}(\bigwedge T^{\ast}M)$.

Let $\{\theta^{\mathbf{a}}\}$, $\theta^{\mathbf{a}}\in\sec\bigwedge^{1}%
T^{\ast}M\hookrightarrow\sec\mathcal{C}\ell(M,\mathtt{g})$, $\mathbf{a=}%
0,1,2,3,$ be an orthonormal basis of $T^{\ast}M$ dual to the basis
$\{\mathbf{e}_{\mathbf{a}}\}$ of $TM$ and let $\omega_{\mathbf{b}}%
^{\mathbf{a}}\in\sec\bigwedge^{1}T^{\ast}M\hookrightarrow\sec\mathcal{C}%
\ell(M,\mathtt{g})$ be the connection $1$-forms of the connection $\nabla$ in
a \textit{given} gauge. We introduce also the 1-jet bundle $J^{1}[(\bigwedge
T^{\ast}M)^{n+2}]$ over the configuration space $(\bigwedge T^{\ast}%
M)^{n+2}\hookrightarrow(\mathcal{C}\ell(M,\mathtt{g}))^{n+2}$ of a field
theory describing $n$ different fields $\phi^{A}\in\sec\bigwedge
T^{p}M\hookrightarrow\sec\mathcal{C}\ell(M,\mathtt{g})$ on a RCST, where for
each different value of $A$ we have in general a different value of $p$.
\begin{align}
J^{1}[(\bigwedge T^{\ast}M)^{n+2}]  &  :=J^{1}(\bigwedge T^{\ast}%
M\times\bigwedge T^{\ast}M\times...\times\bigwedge T^{\ast}M)\nonumber\\
&  =\{(x,\theta^{\mathbf{a}}(x),d\theta^{\mathbf{a}}(x),\omega_{\mathbf{b}%
}^{\mathbf{a}}(x),d\omega_{\mathbf{b}}^{\mathbf{a}}(x),\phi^{A}(x),d\phi
^{A}(x),\text{ }A=1,...,n\}
\end{align}
Sections of $J^{1}[(\bigwedge T^{\ast}M)^{n+2}]$ will be denoted by
$j_{1}(\theta^{\mathbf{a}},\omega_{\mathbf{b}}^{\mathbf{a}},\phi)$ or simply
by $j_{1}(\phi)$ when no confusion arises.

A functional for a field $\phi\in\sec\bigwedge T^{\ast}M\hookrightarrow
\sec\mathcal{C}\ell(M,\mathtt{g})$ in $J^{1}(\bigwedge T^{\ast}M)$ is a
mapping $\mathcal{F}:\sec J^{1}(\bigwedge T^{\ast}M)\rightarrow\sec\bigwedge
T^{\ast}M$, $j_{1}(\phi)\mapsto\mathcal{F}(j_{1}(\phi))$.

A Lagrangian density \textit{mapping} for a field theory described by
$\ $fields $\phi^{A}\in\sec\bigwedge T^{\ast}M$, $A=1,2,...,n$ over a
Riemann-Cartan spacetime is a mapping
\begin{equation}
\mathcal{L}_{m}:\sec J^{1}[(\bigwedge T^{\ast}M)^{n+2}]\rightarrow\sec%
{\displaystyle\bigwedge\nolimits^{4}}
T^{\ast}M,
\end{equation}

\begin{equation}
j_{1}(\theta^{\mathbf{a}},\omega_{\mathbf{b}}^{\mathbf{a}},\phi)\mapsto
\mathcal{L}_{m}(j_{1}(\theta^{\mathbf{a}},\omega_{\mathbf{b}}^{\mathbf{a}%
},\phi)).
\end{equation}
\ 

\begin{remark}
When convenient the image of $\mathcal{L}_{m}$, i.e., $\mathcal{L}_{m}%
(j_{1}(\theta^{\mathbf{a}},\omega_{\mathbf{b}}^{\mathbf{a}},\phi))$
\emph{(}called Lagrangian density\emph{)} will be represented by the sloppy
notation $\mathcal{L}_{m}(x,\theta^{\mathbf{a}},\omega_{\mathbf{b}%
}^{\mathbf{a}},\phi)$ or, when the Lagrangian density does not depend
explicitly on $x$, $\mathcal{L}_{m}(\theta^{\mathbf{a}},\omega_{\mathbf{b}%
}^{\mathbf{a}},\phi)$ or simply $\mathcal{L}_{m}(\phi)$ and even just
$\mathcal{L}_{m}$. The same observation holds for any other functional.
\end{remark}

To simplify the notation even further consider in the next few definitions of
a field theory with only one field $\phi\in\sec\bigwedge^{r}T^{\ast
}M\hookrightarrow\sec\mathcal{C}\ell(M,\mathtt{g})$, in which case
$\mathcal{L}_{m}$ is a functional on $J^{1}[(\bigwedge T^{\ast}M)^{3}]$.

Given a Lagrangian density $\mathcal{L}_{m}(j_{1}(\theta^{\mathbf{a}}%
,\omega_{\mathbf{b}}^{\mathbf{a}},\phi))$ for a given homogeneous matter field
$\phi\in\sec\bigwedge^{r}T^{\ast}M\hookrightarrow\sec\mathcal{C}%
\ell(M,\mathtt{g})$ over a general Riemann-Cartan spacetime, we shall need (in
order to apply the variational action principle) to calculate some algebraic
derivatives of $\mathcal{L}_{m}$. These are terms such as $\frac
{\partial\mathcal{L}_{m}(\phi)}{\partial\phi}$,$\frac{\partial\mathcal{L}%
_{m}(\phi)}{\partial d\phi}$ which appears in the variation of $\mathcal{L}%
_{m}$, i.e.,
\begin{subequations}
\begin{align}%
\mbox{\boldmath{$\delta$}}%
\mathcal{L}_{m}(\phi)  &  =%
\mbox{\boldmath{$\delta$}}%
\phi\wedge\frac{\partial\mathcal{L}_{m}(\phi)}{\partial\phi}+%
\mbox{\boldmath{$\delta$}}%
(d\phi)\wedge\frac{\partial\mathcal{L}_{m}(\phi)}{\partial d\phi}\nonumber\\
&  =%
\mbox{\boldmath{$\delta$}}%
\phi\wedge\frac{\partial\mathcal{L}_{m}(\phi)}{\partial\phi}+d(%
\mbox{\boldmath{$\delta$}}%
\phi)\wedge\frac{\partial\mathcal{L}_{m}(\phi)}{\partial d\phi}\nonumber\\
&  =%
\mbox{\boldmath{$\delta$}}%
\phi\wedge\left(  \frac{\partial\mathcal{L}_{m}(\phi)}{\partial\phi}%
-(-1)^{r}d\left(  \frac{\partial\mathcal{L}_{m}(\phi)}{\partial d\phi}\right)
\right)  +d\left(
\mbox{\boldmath{$\delta$}}%
\phi\wedge\frac{\partial\mathcal{L}_{m}(\phi)}{\partial d\phi}\right)
\nonumber\\
&  =%
\mbox{\boldmath{$\delta$}}%
\phi\wedge\mathcal{\star}\mathbf{\Sigma}\mathcal{(\phi)}+d\left(
\mbox{\boldmath{$\delta$}}%
\phi\wedge\frac{\partial\mathcal{L}_{m}(\phi)}{\partial d\phi}\right)  .
\label{7.1b}%
\end{align}

\end{subequations}
\begin{definition}
The terms $\frac{\partial\mathcal{L}_{m}}{\partial\phi}$ and $\frac
{\partial\mathcal{L}_{m}}{\partial d\phi}$ are called in what follows
algebraic derivatives of $\mathcal{L}_{m}$\ \footnote{This terminology was
originally introduced in \cite{thiwal}. The exterior product \ $%
\mbox{\boldmath{$\delta$}}%
\phi\wedge\frac{\partial}{\partial\phi}$ is a particular instance of the
$A\wedge\frac{\partial}{\partial\phi}$ directional derivatives introduced in
the multiform calculus developed in \cite{rodoliv2006} with $%
\mbox{\boldmath{$\delta$}}%
\phi=A$.} and $\mathcal{\star\Sigma(\phi)}\in\sec\bigwedge^{3}T^{\ast
}M\hookrightarrow\sec\mathcal{C}\ell(M,\mathtt{g})$,
\begin{equation}
\mathcal{\star}\mathbf{\Sigma}\mathcal{(\phi)}=\frac{\partial\mathcal{L}%
_{m}(\phi)}{\partial\phi}-(-1)^{r}d\left(  \frac{\partial\mathcal{L}_{m}%
(\phi)}{\partial d\phi}\right)  \label{7.1bb}%
\end{equation}
is called the Euler-Lagrange functional of the field $\phi$. Some authors call
it the functional derivative of $\mathcal{L}_{m}$ and in this case write%
\begin{equation}
\mathcal{\star}\mathbf{\Sigma}\mathcal{(\phi)=}\frac{%
\mbox{\boldmath{$\delta$}}%
\mathcal{L}_{m}(\phi)}{%
\mbox{\boldmath{$\delta$}}%
\phi} \label{7.1bbb}%
\end{equation}

\end{definition}

In working with these objects it is necessary to keep in mind that for
$\phi\in\sec\bigwedge^{r}T^{\ast}M$, $\mathcal{F}(\phi)\equiv\mathcal{F}%
(j_{1}(\phi))\in\sec\bigwedge^{p}T^{\ast}M$ and $\mathcal{K}(\phi
)\equiv\mathcal{K}(j_{1}(\phi))\in\sec\bigwedge^{q}T^{\ast}M$,%
\begin{equation}
\frac{\partial}{\partial\phi}[\mathcal{F}(\phi)\wedge\mathcal{K}(\phi
)]=\frac{\partial}{\partial\phi}\mathcal{F}(\phi)\wedge\mathcal{K}%
(\phi)+(-1)^{pr}\mathcal{F}(\phi)\wedge\frac{\partial}{\partial\phi
}\mathcal{K}(\phi). \label{7.2}%
\end{equation}

We recall also that if $\mathcal{G(}j_{1}(\phi)\mathcal{)}\in\sec\bigwedge
^{p}T^{\ast}M$ is an arbitrary functional and $\sigma:M\rightarrow M$ a
diffeomorphism, then $\mathcal{G(}j_{1}(\phi)\mathcal{)}$ is said to be
invariant under $\sigma$ if and only if $\sigma^{\ast}\mathcal{G(}%
j_{1}\mathcal{(\phi))=G(}j_{1}\mathcal{(\phi))}$. Also, it is a well known
result that $\mathcal{G(}j_{1}(\phi)\mathcal{)}$ is invariant under the action
of a one parameter group of diffeomorphisms $\sigma_{t}$ if and only if
\begin{equation}
\pounds _{\xi}\mathcal{G(}j_{1}(\phi)\mathcal{)}=0, \label{7.2a}%
\end{equation}
where $\xi\in\sec TM$ is the infinitesimal generator of the group $\sigma_{t}$
and $\pounds _{\xi}$ denotes the Lie derivative.

\subsection{Euler-Lagrange Equations from Lagrangian Densities}

Recall now that the principle of stationary action is the statement that the
variation of the action integral written in terms of a Lagrangian density
$\mathcal{L}_{m}(j_{1}(\theta^{\mathbf{a}},\omega_{\mathbf{b}}^{\mathbf{a}%
},\phi))$ is null for arbitrary variations of $\phi$ which vanish in the
boundary $\partial U$ of the open set $U\subset M$ (i.e., $\left.
\mbox{\boldmath{$\delta$}}%
\phi\right\vert _{\partial U}=0$)
\begin{equation}%
\mbox{\boldmath{$\delta$}}%
\mathcal{A}(\phi)\mathcal{=}%
\mbox{\boldmath{$\delta$}}%
\int\nolimits_{U}\mathcal{L}_{m}(j_{1}(\theta^{\mathbf{a}},\omega_{\mathbf{b}%
}^{\mathbf{a}},\phi))=\int\nolimits_{U}%
\mbox{\boldmath{$\delta$}}%
\mathcal{L}_{m}(j_{1}(\theta^{\mathbf{a}},\omega_{\mathbf{b}}^{\mathbf{a}%
},\phi))=0. \label{7.3}%
\end{equation}

A trivial calculation gives%
\begin{equation}%
\mbox{\boldmath{$\delta$}}%
\mathcal{A}(\phi)=\int\nolimits_{U}%
\mbox{\boldmath{$\delta$}}%
\phi\wedge\star\mathbf{\Sigma}(\phi). \label{7.4}%
\end{equation}
Since $%
\mbox{\boldmath{$\delta$}}%
\phi$ is arbitrary, the stationary action principle implies that
\begin{equation}
\star\mathbf{\Sigma}(\phi)=\frac{\partial\mathcal{L}_{m}(\phi)}{\partial\phi
}-(-1)^{r}d\left(  \frac{\partial\mathcal{L}_{m}(\phi)}{\partial d\phi
}\right)  =0. \label{7.5}%
\end{equation}
The equation $\star\mathbf{\Sigma}(\phi)=0$ is the corresponding \textit{ELE}
for the field $\phi\in\sec\bigwedge^{r}T^{\ast}M\hookrightarrow\sec
\mathcal{C}\ell(M,\mathtt{g})$.

\subsection{Invariance of the Action Integral under the Action of a
Diffeomorphism}

\begin{proposition}
The action $\mathcal{A}(\phi)$ for any field theory formulated in terms of
fields that are differential forms is invariant under the action of one
parameters groups of diffeomorphisms if $\left.  \mathcal{L}_{m}(j_{1}%
(\theta^{\mathbf{a}},\omega_{\mathbf{b}}^{\mathbf{a}},\phi))\right\vert
_{\partial U}=0$ on the boundary $\partial U$ of a domain $U\subset M$.
\end{proposition}

\begin{proof}
Let $\mathcal{L}_{m}(j_{1}(\theta^{\mathbf{a}},\omega_{\mathbf{b}}%
^{\mathbf{a}},\phi))$ be the Lagrangian density of the theory. The variation
of the action which we are interested in is the horizontal variation, i.e.:%
\begin{equation}%
\mbox{\boldmath{$\delta$}}%
_{h}\mathcal{A}(\phi)=\int\nolimits_{U}\pounds _{\xi}\mathcal{L}_{m}%
(j_{1}(\theta^{\mathbf{a}},\omega_{\mathbf{b}}^{\mathbf{a}},\phi)) \label{7.8}%
\end{equation}
Let%
\begin{equation}
\xi^{\ast}=\mathbf{g}(\xi,\cdot)\in\sec\bigwedge\nolimits^{1}T^{\ast
}M\hookrightarrow\sec\mathcal{C}\ell(M,\mathtt{g}). \label{7.8a}%
\end{equation}
Then we have from a well known property of the Lie derivative (Cartan's
magical formula) that
\begin{equation}
\pounds _{\xi}\mathcal{L}_{m}=d(\xi^{\ast}\lrcorner\mathcal{L}_{m})+\xi^{\ast
}\lrcorner(d\mathcal{L}_{m}). \label{7.9}%
\end{equation}
But, since $\mathcal{L}_{m}(j_{1}(\theta^{\mathbf{a}},\omega_{\mathbf{b}%
}^{\mathbf{a}},\phi))\in\sec\bigwedge^{4}T^{\ast}M\hookrightarrow
\sec\mathcal{C}\ell(M,\mathtt{g})$ we have $d\mathcal{L}_{m}=0$ and then
$\pounds _{\xi}\mathcal{L}_{m}=d(\xi^{\ast}\lrcorner\mathcal{L}_{m})$. It
follows, using Stokes theorem that%
\begin{align}
\int\nolimits_{U}\pounds _{\xi}\mathcal{L}_{m}(j_{1}(\theta^{\mathbf{a}%
},\omega_{\mathbf{b}}^{\mathbf{a}},\phi))  &  =\int\nolimits_{U}d[\xi^{\ast
}\lrcorner\mathcal{L}_{m}(j_{1}(\theta^{\mathbf{a}},\omega_{\mathbf{b}%
}^{\mathbf{a}},\phi))]\nonumber\\
&  =\int\nolimits_{\partial U}\xi^{\ast}\lrcorner\mathcal{L}_{m}(j_{1}%
(\theta^{\mathbf{a}},\omega_{\mathbf{b}}^{\mathbf{a}},\phi))=0, \label{7.10}%
\end{align}
since $\left.  \mathcal{L}_{m}(j_{1}(\theta^{\mathbf{a}},\omega_{\mathbf{b}%
}^{\mathbf{a}},\phi))\right\vert _{\partial U}=0$.
\end{proof}

\begin{remark}
It is important to emphasize that the action integral is always invariant
under the action of a one parameter group of diffeomorphisms even if the
corresponding Lagrangian density is not invariant \emph{(}in the sense of
\emph{Eq.(\ref{7.2a}))} under the action of that same group.
\end{remark}

\section{Covariant `Conservation' Laws}

Let $(M,\mathbf{g},\nabla,\tau_{\mathbf{g}},\uparrow)$ denote a general
Riemann-Cartan \textit{spacetime}. As stated above we suppose that the dynamic
fields $\phi^{A}$, $A=1,2,...,n$, are $r$-forms, i.e., each $\phi^{A}\in
\sec\bigwedge^{r}T^{\ast}M\hookrightarrow\sec\mathcal{C}\ell(M,\mathtt{g})$,
for \textit{some} $r=0,1,...,4$.

Let $\{\mathbf{e}_{\mathbf{a}}\}$ be an arbitrary global orthonormal basis for
$TM$, and let $\{\theta^{\mathbf{a}}\}$ be its dual basis. We suppose that
$\theta^{\mathbf{a}}\in\sec\bigwedge^{1}T^{\ast}M\hookrightarrow
\sec\mathcal{C}\ell(M,\mathtt{g})$. Let moreover $\{\theta_{\mathbf{a}}\}$ be
the reciprocal basis of $\{\theta^{\mathbf{a}}\}$. \ As it is well known (see,
e.g., \cite{thiwal,roqui,roqui2005,rodoliv2006}) it is possible to represent
the gravitational field using $\{\theta^{\mathbf{a}}\}$ and it is also
possible to write differential equations equivalent to Einstein equations for
such objects.\footnote{The Lagrangian density for the $\{\theta^{\mathbf{a}%
}\}$ for the case of General Relativity is recalled in Section 5.
\par
{}}

Here, we make the hypothesis that a Riemann-Cartan spacetime models a
generalized gravitational field which must be described by $\{\theta
^{\mathbf{a}},\omega_{\mathbf{b}}^{\mathbf{a}}\}$, where $\omega_{\mathbf{b}%
}^{\mathbf{a}}$ are the connection $1$-forms (in a given gauge). Thus, we
suppose that a dynamic theory for the fields $\phi^{A}\in\sec\bigwedge
^{r}T^{\ast}M$ (called in what follows matter fields) is obtained through the
introduction of a Lagrangian density, which is a functional on $J^{1}%
[(\bigwedge T^{\ast}M)^{2+n}]$ as previously discussed.

Active \textit{Local} Lorentz transformations are represented by \textit{even}
sections of the Clifford bundle $U\in\sec\mathrm{Spin}_{1,3}^{e}%
(M)\hookrightarrow\sec\mathcal{C}\ell^{(0)}(M,\mathtt{g})$, such that
$U\tilde{U}=\tilde{U}U=1$, i.e., $U(x)\in\mathrm{Spin}_{1,3}^{e}%
\simeq\mathrm{Sl}(2,\mathbb{C})$. Under a local Lorentz transformation the
fields transform as
\begin{align}
\theta^{\mathbf{a}}  &  \mapsto\theta^{\prime\mathbf{a}}=U\theta^{\mathbf{a}%
}U^{-1}=\Lambda_{\mathbf{b}}^{\mathbf{a}}\theta^{\mathbf{b}},\nonumber\\
\omega_{\mathbf{b}}^{\mathbf{a}}  &  \mapsto\omega_{\mathbf{b}}^{\prime
\mathbf{a}}=\Lambda_{\mathbf{c}}^{\mathbf{a}}\omega_{\mathbf{d}}^{\mathbf{c}%
}(\Lambda^{-1})_{\mathbf{b}}^{\mathbf{d}}+\Lambda_{\mathbf{c}}^{\mathbf{a}%
}(d\Lambda^{-1})_{\mathbf{b}}^{\mathbf{c}},\label{7.12}\\
\phi^{A}  &  \mapsto\phi^{\prime A}=U\phi^{A}U^{-1},\nonumber
\end{align}
where $\Lambda_{\mathbf{b}}^{\mathbf{a}}(x)\in\mathrm{SO}_{1,3}^{e}$. In our
formalism it is a triviality to see that $\mathcal{L}_{m}(\theta^{\mathbf{a}%
},\omega_{\mathbf{b}}^{\mathbf{a}},\phi)\in\sec\bigwedge\nolimits^{4}T^{\ast
}M\hookrightarrow\sec\mathcal{C}\ell(M,\mathtt{g})$ is invariant under local
Lorentz transformations. Indeed, since $\tau_{\mathtt{g}}=\theta^{\mathbf{5}%
}=\theta^{\mathbf{0}}\theta^{\mathbf{1}}\theta^{\mathbf{2}}\theta^{\mathbf{3}%
}$ $\in\sec\bigwedge\nolimits^{4}T^{\ast}M\hookrightarrow\sec\mathcal{C}%
\ell(M,\mathtt{g})$ commutes with even multiform fields, we have that a local
Lorentz transformation produces no changes in $\mathcal{L}_{m}$, i.e.,
\begin{equation}
\mathcal{L}_{m}(\theta^{\mathbf{a}},\omega_{\mathbf{b}}^{\mathbf{a}}%
,\phi)\mapsto U\mathcal{L}_{m}(\theta^{\mathbf{a}},\omega_{\mathbf{b}%
}^{\mathbf{a}},\phi)U^{-1}=\mathcal{L}_{m}(\theta^{\mathbf{a}},\omega
_{\mathbf{b}}^{\mathbf{a}},\phi). \label{7.12a}%
\end{equation}
However, this does not implies necessarily that the variation of the
Lagrangian density $\mathcal{L}_{m}(\theta^{\mathbf{a}},\omega_{\mathbf{b}%
}^{\mathbf{a}},\phi)$ obtained by variation of the fields $(\theta
^{\mathbf{a}},\omega_{\mathbf{b}}^{\mathbf{a}},\phi)$ is null, since%
\begin{equation}%
\mbox{\boldmath{$\delta$}}%
_{v}\mathcal{L}_{m}=\mathcal{L}_{m}(\theta^{\mathbf{a}}+%
\mbox{\boldmath{$\delta$}}%
_{v}\theta^{\mathbf{a}},\omega_{\mathbf{b}}^{\mathbf{a}}+%
\mbox{\boldmath{$\delta$}}%
_{v}\omega_{\mathbf{b}}^{\mathbf{a}},\phi+%
\mbox{\boldmath{$\delta$}}%
_{v}\phi)-\mathcal{L}_{m}(\theta^{\mathbf{a}},\omega_{\mathbf{b}}^{\mathbf{a}%
},\phi)\neq0,
\end{equation}
unless it happens that for an infinitesimal Lorentz transformation,%
\begin{align}
&  \mathcal{L}_{m}(\theta^{\mathbf{a}}+%
\mbox{\boldmath{$\delta$}}%
_{v}\theta^{\mathbf{a}},\omega_{\mathbf{b}}^{\mathbf{a}}+%
\mbox{\boldmath{$\delta$}}%
_{v}\omega_{\mathbf{b}}^{\mathbf{a}},\phi+%
\mbox{\boldmath{$\delta$}}%
_{v}\phi)\nonumber\\
&  =\mathcal{L}_{m}(U\theta^{\mathbf{a}}U^{-1},U\omega_{\mathbf{b}%
}^{\mathbf{a}}U^{-1},U\phi U^{-1})=U\mathcal{L}_{m}U^{-1}=\mathcal{L}_{m}.
\end{align}

In what follows we suppose that the Lagrangian of the matter field is
invariant under local Lorentz transformations\footnote{We discuss further the
issue of local Lorentz invariance and its hidden consequence in
\cite{rodroldao,roldaowa}.}, i.e., $%
\mbox{\boldmath{$\delta$}}%
_{v}\mathcal{L}_{m}=0$. \ Also $\mathcal{L}_{m}$ depends on the $\theta
^{\mathbf{a}}$ and $\omega_{\mathbf{b}}^{\mathbf{a}}$,\ but not on
$d\theta^{\mathbf{a}}$ and $d\omega_{\mathbf{b}}^{\mathbf{a}}$ (minimal
coupling).\footnote{See example of the electromagnetic field in Appendice B.}
Then, $\frac{%
\mbox{\boldmath{$\delta$}}%
\mathcal{L}_{m}}{%
\mbox{\boldmath{$\delta$}}%
\theta^{\mathbf{a}}}=\frac{\partial\mathcal{L}_{m}}{\partial\theta
^{\mathbf{a}}}$ and $\frac{%
\mbox{\boldmath{$\delta$}}%
\mathcal{L}_{m}}{%
\mbox{\boldmath{$\delta$}}%
\omega_{\mathbf{b}}^{\mathbf{a}}}=\frac{\partial\mathcal{L}_{m}}%
{\partial\omega_{\mathbf{b}}^{\mathbf{a}}}$ and we can write
\begin{equation}
\int%
\mbox{\boldmath{$\delta$}}%
\mathcal{L}_{m}=\int\left[
\mbox{\boldmath{$\delta$}}%
\theta^{\mathbf{a}}\wedge\frac{\partial\mathcal{L}_{m}}{\partial
\theta^{\mathbf{a}}}+%
\mbox{\boldmath{$\delta$}}%
\omega_{\mathbf{b}}^{\mathbf{a}}\wedge\frac{\partial\mathcal{L}_{m}}%
{\partial\omega_{\mathbf{b}}^{\mathbf{a}}}+%
\mbox{\boldmath{$\delta$}}%
\phi^{A}\wedge\mathop\star\mathbf{\Sigma}_{A}\right]  ,\label{7.14}%
\end{equation}
where $\mathbf{\Sigma}_{A}$ \ are the Euler-Lagrange functionals of the fields
$\phi^{A}.$

As we just showed above the action of any Lagrangian density is invariant
under diffeomorphisms. Let us now calculate the total variation of the
Lagrangian density $\mathcal{L}_{m}$, arising from a one-parameter group of
diffeomorphisms generated by a vector field $\xi\in\sec TM$ and by a local
Lorentz transformation, when we vary $\theta^{\mathbf{a}},\omega_{\mathbf{b}%
}^{\mathbf{a}},\phi^{A},d\phi^{A}$ independently. We have
\begin{equation}%
\mbox{\boldmath{$\delta$}}%
\mathcal{L}_{m}=%
\mbox{\boldmath{$\delta$}}%
_{v}\mathcal{L}_{m}-\pounds _{\xi}\mathcal{L}_{m}. \label{7.13}%
\end{equation}

Under the (nontrivial) hypothesis \cite{rodroldao,roldaowa} that $%
\mbox{\boldmath{$\delta$}}%
_{v}\mathcal{L}_{m}=0$,%
\begin{equation}%
\mbox{\boldmath{$\delta$}}%
\mathcal{L}_{m}=-\pounds _{\xi}\mathcal{L}_{m}=-\star\mathcal{T}_{\mathbf{a}%
}\wedge\pounds _{\xi}\theta^{\mathbf{a}}-\star\mathbf{J}_{\mathbf{a}%
}^{\mathbf{b}}\wedge\pounds _{\xi}\omega_{\mathbf{b}}^{\mathbf{a}}%
-\star\mathbf{\Sigma}_{A}\wedge\pounds _{\xi}\phi^{A}, \label{7.17}%
\end{equation}
where we have:

\begin{definition}
The coefficients of $%
\mbox{\boldmath{$\delta$}}%
\theta^{\mathbf{a}}=-\pounds _{\xi}\theta^{\mathbf{a}}$, i.e.
\begin{equation}
\star\mathcal{T}_{a}=\frac{\partial\mathcal{L}_{m}}{\partial\theta
^{\mathbf{a}}}\in\sec\bigwedge\nolimits^{3}T^{\ast}M \label{7.15}%
\end{equation}
are called the energy-momentum densities of the matter fields, and the
\ $\mathcal{T}_{\mathbf{a}}\in\sec\bigwedge^{1}T^{\ast}M$ are called the
energy-momentum density $1$-forms of the matter fields. The coefficients of $%
\mbox{\boldmath{$\delta$}}%
\omega_{\mathbf{b}}^{\mathbf{a}}$, i.e.,%
\begin{equation}
\star\mathbf{J}_{\mathbf{a}}^{\mathbf{b}}=\frac{\partial\mathcal{L}_{m}%
}{\partial\omega_{\mathbf{b}}^{\mathbf{a}}}\in\sec\bigwedge\nolimits^{3}%
T^{\ast}M, \label{7.16}%
\end{equation}
are called the angular momentum densities of the matter fields.
\end{definition}

Taking into account that each one of the fields $\phi^{A}$ obey a
Euler-Lagrange equation, $\mathop\star\mathbf{\Sigma}_{A}=0$, we can write%
\begin{equation}
\int\pounds _{\xi}\mathcal{L}_{m}=\int\star\mathcal{T}_{\mathbf{a}}%
\wedge\pounds _{\xi}\theta^{\mathbf{a}}+\star\mathbf{J}_{\mathbf{a}%
}^{\mathbf{b}}\wedge\pounds _{\xi}\omega_{\mathbf{b}}^{\mathbf{a}}
\label{7.18}%
\end{equation}

Now, since all geometrical objects in the above formulas are sections of the
Clifford bundle, we can write
\begin{equation}
\pounds _{\xi}\theta^{\mathbf{a}}=\xi^{\ast}\lrcorner d\theta^{\mathbf{a}%
}+d(\xi^{\ast}\lrcorner\theta^{\mathbf{a}}). \label{7.19}%
\end{equation}
Moreover, recalling also Cartan's first structure equation,%
\begin{equation}
d\theta^{\mathbf{a}}+\omega_{\mathbf{b}}^{\mathbf{a}}\wedge\theta^{\mathbf{b}%
}=\Theta^{\mathbf{a}}, \label{7.20}%
\end{equation}
we get%
\begin{align}
\pounds _{\xi}\theta^{\mathbf{a}}  &  =\xi^{\ast}\lrcorner\Theta^{\mathbf{a}%
}-\xi^{\ast}\lrcorner\left(  \omega_{\mathbf{b}}^{\mathbf{a}}\wedge
\theta^{\mathbf{b}}\right)  +d(\xi^{\ast}\lrcorner\theta^{\mathbf{a}%
})\nonumber\\
&  =\xi^{\ast}\lrcorner\Theta^{\mathbf{a}}-(\xi^{\ast}\cdot\omega_{\mathbf{b}%
}^{\mathbf{a}})\theta^{\mathbf{b}}+\left(  \xi^{\ast}\cdot\theta^{\mathbf{b}%
}\right)  \omega_{\mathbf{b}}^{\mathbf{a}}+d(\xi^{\ast}\lrcorner
\theta^{\mathbf{a}})\nonumber\\
&  =\mathbf{D}(\xi^{\ast}\lrcorner\theta^{\mathbf{a}})+\xi^{\ast}%
\lrcorner\Theta^{\mathbf{a}}-(\xi^{\ast}\cdot\omega_{\mathbf{b}}^{\mathbf{a}%
})\theta^{\mathbf{b}}, \label{7.21}%
\end{align}
where $\mathbf{D}$\textbf{ }is the covariant exterior derivative of indexed
$p$-form fields (for details, see, e.g., \cite{benntucker,rodoliv2006}). To
continue we need the following

\begin{proposition}
\label{liealgebrat}Let $\mathbf{\omega}$ be the $4\times4$ matrix whose
entries are the connection $1$-forms. For any $x\in M$, the matrix with
entries $\xi^{\ast}\lrcorner\omega_{\mathbf{b}}^{\mathbf{a}}\in\mathrm{spin}%
_{1,3}^{e}\simeq\mathrm{sl}(2,\mathbb{C})=\mathrm{so}_{1,3}^{e}$ belongs to
the Lie algebra of \textrm{Spin}$_{1,3}^{e}$ \emph{(}or of $\mathrm{SO}%
_{1,3}^{e}$\emph{)}.
\end{proposition}

\begin{proof}
Recall that at any $x\in M$ any infinitesimal local Lorentz transformation
$\ \Lambda_{\mathbf{b}}^{\mathbf{a}}(x)\in\mathrm{SO}_{1,3}^{e}$ can be
written as%
\begin{align}
\Lambda_{\mathbf{b}}^{\mathbf{a}}  &  =\delta_{\mathbf{b}}^{\mathbf{a}}%
+\chi_{\mathbf{b}}^{\mathbf{a}}\text{, }\;\;\;\;\;\left\vert \chi_{\mathbf{b}%
}^{\mathbf{a}}\right\vert \ll1,\nonumber\\
\chi_{\mathbf{ab}}  &  =-\chi_{\mathbf{ba}}. \label{7.22}%
\end{align}
Now, writing $\omega_{\mathbf{b}}^{\mathbf{a}}=L_{\mathbf{cb}}^{\mathbf{a}%
}\theta^{\mathbf{c}}$ we have
\begin{align}
\xi^{\ast}\cdot\omega_{\mathbf{b}}^{\mathbf{a}}  &  =\xi^{\ast}\cdot
(L_{\mathbf{cb}}^{\mathbf{a}}\theta^{\mathbf{c}})=(\xi_{\mathbf{d}}%
\theta^{\mathbf{d}})\cdot(L_{\mathbf{cb}}^{\mathbf{a}}\theta^{\mathbf{c}%
})\nonumber\\
&  =\xi^{\mathbf{c}}L_{\mathbf{cb}}^{\mathbf{a}} \label{7.23}%
\end{align}
and the\ $\xi^{\ast}\cdot\omega_{\mathbf{ab}}$ satisfy%
\begin{equation}
\xi^{\ast}\cdot\omega_{\mathbf{ab}}+\xi^{\ast}\cdot\omega_{\mathbf{ba}}%
=\xi^{\mathbf{c}}(L_{\mathbf{acb}}+L_{\mathbf{bca}})=0, \label{7.24}%
\end{equation}
since in an orthonormal basis the connection coefficients satisfy
$L_{\mathbf{acb}}=-L_{\mathbf{bca}}$. We see then that we can identify if
$\left\vert \xi^{\mathbf{c}}\right\vert \ll1$
\begin{equation}
\chi_{\mathbf{b}}^{\mathbf{a}}=\xi^{\ast}\cdot\omega_{\mathbf{b}}^{\mathbf{a}}
\label{7.25}%
\end{equation}
as the generator of an infinitesimal Lorentz transformation, and the
proposition is proved.
\end{proof}

Now, the term $\left(  \xi^{\ast}\cdot\omega_{\mathbf{b}}^{\mathbf{a}}\right)
\theta^{\mathbf{b}}$ has the form of a local \textit{vertical} variation of
the $\theta^{\mathbf{a}}$ and thus we write
\begin{equation}%
\mbox{\boldmath{$\delta$}}%
_{v}\theta^{\mathbf{a}}:=\left(  \xi^{\ast}\cdot\omega_{\mathbf{b}%
}^{\mathbf{a}}\right)  \theta^{\mathbf{b}} \label{7.26}%
\end{equation}
Using Eq.(\ref{7.26}) we can rewrite Eq.(\ref{7.21}) as%
\begin{equation}
\pounds _{\xi}\theta^{\mathbf{a}}=\mathbf{D}(\xi^{\ast}\cdot\theta
^{\mathbf{a}})+\xi^{\ast}\lrcorner\Theta^{\mathbf{a}}-%
\mbox{\boldmath{$\delta$}}%
_{v}\theta^{\mathbf{a}}. \label{7.27}%
\end{equation}

We see that $\pounds _{\xi}\theta^{\mathbf{a}}=-%
\mbox{\boldmath{$\delta$}}%
_{v}\theta^{\mathbf{a}}$ only if we have the following constraint%
\begin{equation}
\mathbf{D}(\xi^{\ast}\cdot\theta^{\mathbf{a}})+\xi^{\ast}\lrcorner
\Theta^{\mathbf{a}}=0. \label{7.27a}%
\end{equation}

A necessary and sufficient condition for the validity of Eq.(\ref{7.27a}) is
given by Lemma \ref{step1} \ below.

Now, let us calculate $\pounds _{\xi}\omega_{\mathbf{b}}^{\mathbf{a}}$. By
definition,
\begin{align}
\pounds _{\xi}\omega_{\mathbf{b}}^{\mathbf{a}}  &  =\xi^{\ast}\lrcorner
(d\omega_{\mathbf{b}}^{\mathbf{a}})+d(\xi^{\ast}\cdot\omega_{\mathbf{b}%
}^{\mathbf{a}})\nonumber\\
&  =\xi^{\ast}\lrcorner(\mathcal{R}_{\mathbf{b}}^{\mathbf{a}})-(\xi^{\ast
}\cdot\omega_{\mathbf{c}}^{\mathbf{a}})\omega_{\mathbf{b}}^{\mathbf{c}}%
+(\xi^{\ast}\cdot\omega_{\mathbf{b}}^{\mathbf{c}})\omega_{\mathbf{c}%
}^{\mathbf{a}}+d(\xi^{\ast}\cdot\omega_{\mathbf{b}}^{\mathbf{a}}),
\label{7.28}%
\end{align}
where in writing the second line in Eq.(\ref{7.28}) we used Cartan's second
structure equation,
\begin{equation}
d\omega_{\mathbf{b}}^{\mathbf{a}}+\omega_{\mathbf{c}}^{\mathbf{a}}\wedge
\omega_{\mathbf{b}}^{\mathbf{c}}=\mathcal{R}_{\mathbf{b}}^{\mathbf{a}}.
\label{7.29}%
\end{equation}
Under an infinitesimal Lorentz transformation $\Lambda=1+\chi$, recalling
Eq.(\ref{7.12}), we can write
\begin{equation}%
\mbox{\boldmath{$\delta$}}%
_{v}\omega=-d\chi+\chi\omega-\omega\chi, \label{7.30}%
\end{equation}
which using Eq.(\ref{7.25}) gives for Eq.(\ref{7.28})%
\begin{equation}
\pounds _{\xi}\omega_{\mathbf{b}}^{\mathbf{a}}=\xi^{\ast}\lrcorner
(\mathcal{R}_{\mathbf{b}}^{\mathbf{a}})-%
\mbox{\boldmath{$\delta$}}%
_{v}\omega_{\mathbf{b}}^{\mathbf{a}} \label{7.31}%
\end{equation}

Now, for a vertical variation,%

\begin{equation}%
{\displaystyle\int\nolimits_{U}}
\mbox{\boldmath{$\delta$}}%
_{v}\mathcal{L}_{m}:=%
{\displaystyle\int\nolimits_{U}}
\mbox{\boldmath{$\delta$}}%
_{v}\theta^{\mathbf{a}}\wedge\frac{\partial\mathcal{L}_{m}}{\partial
\theta^{\mathbf{a}}}+%
\mbox{\boldmath{$\delta$}}%
_{v}\omega_{\mathbf{b}}^{\mathbf{a}}\wedge\frac{\partial\mathcal{L}_{m}%
}{\partial\omega_{\mathbf{b}}^{\mathbf{a}}}+%
\mbox{\boldmath{$\delta$}}%
_{v}\phi^{A}\wedge\frac{%
\mbox{\boldmath{$\delta$}}%
\mathcal{L}_{m}}{%
\mbox{\boldmath{$\delta$}}%
\phi^{A}}. \label{7.31bis}%
\end{equation}
Then, if we recall that we assumed that $\int%
\mbox{\boldmath{$\delta$}}%
_{v}\mathcal{L}_{m}=0$ and if we suppose that the field equations are
satisfied, i.e., $\star\mathbf{\Sigma}_{A}=\frac{%
\mbox{\boldmath{$\delta$}}%
\mathcal{L}_{m}}{%
\mbox{\boldmath{$\delta$}}%
\phi^{A}}=0$, Eq.(\ref{7.18}) becomes,
\begin{align}
&  \int\pounds _{\xi}\mathcal{L}_{m}\nonumber\\
&  =\int\left[  -\mathbf{D(}\xi^{\ast}\cdot\theta^{\mathbf{a}}\mathbf{)-(}%
\xi^{\ast}\lrcorner\Theta^{\mathbf{a}}\mathbf{)+}%
\mbox{\boldmath{$\delta$}}%
_{v}\theta^{\mathbf{a}}\right]  \wedge\star\mathcal{T}_{\mathbf{a}}\nonumber\\
&  +\int\left[  -\xi^{\ast}\lrcorner(\mathcal{R}_{\mathbf{b}}^{\mathbf{a}})+%
\mbox{\boldmath{$\delta$}}%
_{v}\omega_{\mathbf{b}}^{\mathbf{a}}\right]  \wedge\star\mathbf{J}%
_{\mathbf{a}}^{\mathbf{b}}\nonumber\\
&  =\int\star\mathcal{T}_{\mathbf{a}}\wedge\mathbf{(\xi^{\ast}\lrcorner}%
\Theta\mathbf{^{\mathbf{a}})}+\star\mathbf{\mathbf{J}_{\mathbf{a}}%
^{\mathbf{b}}\wedge(}\xi^{\ast}\lrcorner(\mathcal{R}_{\mathbf{b}}^{\mathbf{a}%
})-D[\star\mathcal{T}\mathbf{_{\mathbf{a}}(}\xi^{\ast}\cdot\theta^{\mathbf{a}%
})]+(\mathbf{D}\star\mathcal{T}\mathbf{\mathbf{_{\mathbf{a}})(}}\xi^{\ast
}\mathbf{\cdot}\theta\mathbf{^{\mathbf{a}}\mathbf{)})}\label{7.34}\\
&  =\int\star\mathcal{T}_{\mathbf{a}}\wedge\mathbf{(\xi^{\ast}\lrcorner}%
\Theta\mathbf{^{\mathbf{a}})}+\star\mathbf{\mathbf{J}_{\mathbf{a}}%
^{\mathbf{b}}\wedge(}\xi^{\ast}\lrcorner(\mathcal{R}_{\mathbf{b}}^{\mathbf{a}%
})+(\mathbf{D}\star\mathcal{T}\mathbf{\mathbf{_{\mathbf{a}})(}}\xi^{\ast
}\mathbf{\cdot}\theta\mathbf{^{\mathbf{a}}\mathbf{)}),}%
\end{align}
where we used also the fact that $\mathbf{D[(}\xi^{\ast}\cdot\theta
^{\mathbf{a}})\star\mathcal{T}\mathbf{_{\mathbf{a}}]}$ $=d\mathbf{[(}\xi
^{\ast}\cdot\theta^{\mathbf{a}})\star\mathcal{T}\mathbf{_{\mathbf{a}}]},$ that
$\left.  \star\mathcal{T}\mathbf{_{\mathbf{a}}}\right\vert _{\partial U}=0$
and
\begin{equation}
\int\nolimits_{U}d\mathbf{[(}\xi^{\ast}\cdot\theta^{\mathbf{a}})\star
\mathcal{T}\mathbf{_{\mathbf{a}}]=}\int\nolimits_{\partial U}\mathbf{(}%
\xi^{\ast}\cdot\theta^{\mathbf{a}})\star\mathcal{T}\mathbf{_{\mathbf{a}}=0}
\label{7.35}%
\end{equation}

Now, writing $\xi^{\ast}=\xi^{\mathbf{a}}\theta_{\mathbf{a}}=\xi_{\mathbf{a}%
}\theta^{\mathbf{a}}$, and recalling that the action is invariant under
diffeomorphisms (if as usual we suppose that $\left.  \mathcal{L}%
_{m}\right\vert _{\partial U}=0$), we have,
\begin{equation}
\int-%
\mbox{\boldmath{$\delta$}}%
\mathcal{L}_{m}=\int\pounds _{\xi}\mathcal{L}_{m}=\left[  \star\mathcal{T}%
\mathbf{_{\mathbf{a}}\wedge}\left(  \theta_{\mathbf{c}}\lrcorner
\Theta^{\mathbf{a}}\right)  +\star\mathbf{J}_{\mathbf{b}}^{\mathbf{a}}%
\wedge(\theta_{\mathbf{c}}\lrcorner\mathcal{R}_{\mathbf{b}}^{\mathbf{a}%
})+\mathbf{\mathbf{D}\star}\mathcal{T}\mathbf{\mathbf{_{\mathbf{c}}}}\right]
\xi^{\mathbf{c}}=0, \label{7.36}%
\end{equation}
and since the $\xi^{\mathbf{c}}$ are arbitrary, we end with
\begin{equation}
\mathbf{\mathbf{D}\star}\mathcal{T}_{\mathbf{c}}+\star\mathcal{T}_{\mathbf{a}%
}\wedge\left(  \theta_{\mathbf{c}}\lrcorner\Theta^{\mathbf{a}}\right)
+\star\mathbf{J}_{\mathbf{b}}^{\mathbf{a}}\wedge(\theta_{\mathbf{c}}%
\lrcorner\mathcal{R}_{\mathbf{b}}^{\mathbf{a}})=0. \label{7.37}%
\end{equation}

Also, using the explicit expressions for $%
\mbox{\boldmath{$\delta$}}%
_{v}\theta^{\mathbf{a}}$ and $%
\mbox{\boldmath{$\delta$}}%
_{v}\omega_{\mathbf{b}}^{\mathbf{a}}$ (Eq.(\ref{7.26}) and Eq.(\ref{7.31})) in
Eq.(\ref{7.31bis}) we get,%
\begin{align}
&  \int\star\mathcal{T}\mathbf{_{\mathbf{a}}\wedge}\chi_{\mathbf{b}%
}^{\mathbf{a}}\theta^{\mathbf{b}}+\star\mathbf{J}_{\mathbf{a}}^{\mathbf{b}%
}\wedge\left(  \chi_{\mathbf{c}}^{\mathbf{a}}\omega_{\mathbf{b}}^{\mathbf{c}%
}-\omega_{\mathbf{c}}^{\mathbf{a}}\chi_{\mathbf{b}}^{\mathbf{c}}%
-d\chi_{\mathbf{b}}^{\mathbf{a}}\right) \nonumber\\
&  =\int\left[  \frac{1}{2}\left(  \star\mathcal{T}\mathbf{_{\mathbf{a}}%
\wedge}\theta^{\mathbf{b}}-\star\mathcal{T}^{\mathbf{b}}\mathbf{\wedge}%
\theta_{\mathbf{a}}\right)  -d\mathop\star\mathbf{J}_{\mathbf{a}}^{\mathbf{b}%
}-\omega_{\mathbf{b}}^{\mathbf{c}}\wedge\star\mathbf{J}_{\mathbf{c}%
}^{\mathbf{b}}-\star\mathbf{J}_{\mathbf{a}}^{\mathbf{c}}\wedge\omega
_{\mathbf{c}}^{\mathbf{b}}\right]  \chi_{\mathbf{b}}^{\mathbf{a}}=0,
\label{7.38}%
\end{align}
and since the coefficients $\chi_{\mathbf{b}}^{\mathbf{a}}$ are arbitrary we
end with%
\begin{equation}
\mathbf{D}\star\mathbf{J}_{\mathbf{a}}^{\mathbf{b}}+\frac{1}{2}\left(
\star\mathcal{T}^{\mathbf{b}}\mathbf{\wedge}\theta_{\mathbf{a}}-\star
\mathcal{T}\mathbf{_{\mathbf{a}}\wedge}\theta^{\mathbf{b}}\right)  =0.
\label{7.39}%
\end{equation}

Eq.(\ref{7.37}) and Eq.(\ref{7.39}) are known as \textit{covariant
conservation laws }and first appeared in this form in \cite{benn}. They are
simply identities that follows from the hypothesis utilized, namely that the
Lagrangian density of the theory is invariant under diffeomorphisms and also
invariant under the\textit{ local} action of the group $\mathrm{Spin}%
_{1,3}^{e}$. Eq.(\ref{7.37}) and Eq.(\ref{7.39}) do \textit{not} encode
genuine conservation laws and a memorable number of nonsense affirmations have
been generated along the years by authors that use those equations in a naive
way. Some examples of the these affirmations are recalled in the specific case
of Einstein's theory in Section~5 \cite{roqui2005}.

\section{When Genuine Conservation Laws Do Exist?}

We recall now the crucial result that when the Riemann-Cartan
\textit{spacetime }$(M,\mathbf{g},\nabla,\tau_{\mathbf{g}},\uparrow)$ admits
symmetries, then Eq.(\ref{7.37}) and Eq.(\ref{7.39}) can be used, as first
shown by Trautman \cite{trautman1,trautman2,trautman3,trautman4}, for the
construction of closed $3$-forms, which then provides genuine conservation
laws involving only the energy-momentum and angular momentum tensors of the
matter fields. In the remaining of the section we recall these results
following \cite{benn}.

\begin{proposition}
\label{killing propos}For each Killing vector field $\xi\in\sec TM$, such that
$\pounds _{\xi}\mathtt{\mathbf{g}}=0$ and $\pounds _{\xi}\Theta=0$, where
$\Theta=\mathbf{e}_{\mathbf{a}}\otimes\Theta^{\mathbf{a}}$ is the
\emph{torsion} tensor of $\ \nabla$, and $\Theta^{\mathbf{a}}$ the torsion
$2$-forms, we have
\begin{equation}
d\left[  \left(  \xi^{\ast}\cdot\theta^{\mathbf{a}}\right)  \mathop\star
\mathcal{T}_{\mathbf{a}}+\left(  \theta_{\mathbf{b}}\cdot\mathbf{L}_{\xi
}\theta^{\mathbf{a}}\right)  \mathop\star\mathbf{J}_{\mathbf{a}}^{\mathbf{b}%
}\right]  =0, \label{7.40}%
\end{equation}
where $\mathbf{L}_{\xi}=\xi^{\ast}\mathbin\lrcorner\mathbf{D}+\mathbf{Di}%
_{\xi}$ is the so called Lie covariant derivative.
\end{proposition}

In order to prove the Proposition \ref{killing propos}, some preliminary
results are needed.

\begin{lemma}
\label{step1}$\pounds _{\xi}\theta^{\mathbf{a}}=-%
\mbox{\boldmath{$\delta$}}%
_{v}\theta^{\mathbf{a}}$ and $\pounds _{\xi}\omega_{\mathbf{b}}^{\mathbf{a}}=%
\mbox{\boldmath{$\delta$}}%
_{v}\omega_{\mathbf{b}}^{\mathbf{a}}$ if and only if $\pounds _{\xi
}\mathtt{\mathbf{g}}=0$ and $\pounds _{\xi}\Theta=0$.
\end{lemma}

\begin{proof}
Let us show first that if $\pounds _{\xi}\theta^{\mathbf{a}}=-%
\mbox{\boldmath{$\delta$}}%
_{v}\theta^{\mathbf{a}}$ then $\pounds _{\xi}\mathbf{g}=0$. We have%
\begin{equation}
\pounds _{\xi}\mathbf{g}=\eta_{\mathbf{ab}}\left(  \pounds _{\xi}%
\theta^{\mathbf{a}}\right)  \otimes\theta^{\mathbf{b}}+\eta_{\mathbf{ab}%
}\theta^{\mathbf{a}}\otimes(\pounds _{\xi}\theta^{\mathbf{b}}). \label{7.41}%
\end{equation}
On the other hand, since $\mathtt{\mathbf{g}}$ is invariant under local
Lorentz transformations, we have
\begin{equation}%
\mbox{\boldmath{$\delta$}}%
_{v}\mathbf{g}=\eta_{\mathbf{ab}}\left(
\mbox{\boldmath{$\delta$}}%
_{v}\theta^{\mathbf{a}}\right)  \otimes\theta^{\mathbf{b}}+\eta_{\mathbf{ab}%
}\theta^{\mathbf{a}}\otimes(%
\mbox{\boldmath{$\delta$}}%
_{v}\theta^{\mathbf{b}})=0. \label{7.42}%
\end{equation}
Then, it follows from Eqs.(\ref{7.41}) and (\ref{7.42}) that if $\pounds _{\xi
}\theta^{\mathbf{a}}=-%
\mbox{\boldmath{$\delta$}}%
_{v}\theta^{\mathbf{a}}$ then $\pounds _{\xi}\mathbf{g}=0$.

Taking into account the definition of Lie derivative we can write%
\begin{align}
\pounds _{\xi}\mathbf{e}_{\mathbf{a}}  &  =-\varkappa_{\mathbf{a}}%
^{\mathbf{b}}\mathbf{e}_{\mathbf{b}},\text{ }\pounds _{\xi}\theta^{\mathbf{a}%
}=\varkappa_{\mathbf{b}}^{\mathbf{a}}\theta^{\mathbf{b}},\nonumber\\
\varkappa_{\mathbf{a}}^{\mathbf{b}}  &  =-[\mathbf{e}_{\mathbf{a}}%
(\xi^{\mathbf{b}})+\xi^{\mathbf{m}}c_{\mathbf{am}}^{\mathbf{b}}] \label{7.42a}%
\end{align}

Now, if $\pounds _{\xi}\mathbf{g}=0$ we have from Eq.(\ref{7.41}) that
$\left(  \eta_{\mathbf{cb}}\varkappa_{\mathbf{a}}^{\mathbf{c}}+\eta
_{\mathbf{ac}}\varkappa_{\mathbf{b}}^{\mathbf{c}}\right)  \theta^{\mathbf{a}%
}\otimes\theta^{\mathbf{b}}=0$, i.e.,%
\begin{equation}
\varkappa_{\mathbf{ab}}+\varkappa_{\mathbf{ba}}=0, \label{7.43}%
\end{equation}
and then it follows that for any $x\in M$, $\varkappa_{\mathbf{ab}}%
\in\mathrm{spin}_{1,3}^{e}$. Using Proposition \ref{liealgebrat} we can write
$\varkappa_{\mathbf{b}}^{\mathbf{a}}=-\chi_{\mathbf{b}}^{\mathbf{a}}%
=-\xi^{\ast}\cdot\omega_{\mathbf{b}}^{\mathbf{a}}$ and then the vertical
variation can be written as $%
\mbox{\boldmath{$\delta$}}%
_{v}\theta^{\mathbf{a}}=-\pounds _{\xi}\theta^{\mathbf{a}}$ .

The proof that if $\pounds _{\xi}\omega_{\mathbf{b}}^{\mathbf{a}}=%
\mbox{\boldmath{$\delta$}}%
_{v}\omega_{\mathbf{b}}^{\mathbf{a}}$ then $\pounds _{\xi}\Theta=0$ is
trivial. In the following we prove the reciprocal, i.e., if $\pounds _{\xi
}\Theta=0$ then $\pounds _{\xi}\omega_{\mathbf{b}}^{\mathbf{a}}=%
\mbox{\boldmath{$\delta$}}%
_{v}\omega_{\mathbf{b}}^{\mathbf{a}}$. \ We have,%
\begin{equation}
\pounds _{\xi}\Theta=\pounds _{\xi}\mathbf{e}_{\mathbf{a}}\otimes
\Theta^{\mathbf{a}}+\mathbf{e}_{\mathbf{a}}\otimes\pounds _{\xi}%
\Theta^{\mathbf{a}} \label{7.44}%
\end{equation}
Then, if $\pounds _{\xi}\Theta=0$ we conclude that%
\begin{equation}
\pounds _{\xi}\Theta^{\mathbf{a}}=\varkappa_{\mathbf{b}}^{\mathbf{a}}%
\Theta^{\mathbf{b}}, \label{7.45}%
\end{equation}
which is an infinitesimal Lorentz transformation of the torsion $2$-forms. On
the other hand, taking into account Cartan's first structure equation,
Eq.(\ref{7.42a}), and the fact that $\pounds _{\xi}d\theta^{\mathbf{a}%
}=d(\pounds _{\xi}\theta^{\mathbf{a}})$, we can write%
\begin{align}
\pounds _{\xi}\Theta^{\mathbf{a}}  &  =\pounds _{\xi}d\theta^{\mathbf{a}%
}+\pounds _{\xi}\omega_{\mathbf{b}}^{\mathbf{a}}\wedge\theta^{\mathbf{b}%
}+\omega_{\mathbf{b}}^{\mathbf{a}}\wedge\pounds _{\xi}\theta^{\mathbf{b}%
}\nonumber\\
&  =d\left(  \varkappa_{\mathbf{b}}^{\mathbf{a}}\theta^{\mathbf{b}}\right)
+\pounds _{\xi}\omega_{\mathbf{b}}^{\mathbf{a}}\wedge\theta^{\mathbf{b}%
}+\omega_{\mathbf{b}}^{\mathbf{a}}\wedge\varkappa_{\mathbf{c}}^{\mathbf{b}%
}\theta^{\mathbf{c}}\nonumber\\
&  =d\left(  \varkappa_{\mathbf{b}}^{\mathbf{a}}\right)  \wedge\theta
^{\mathbf{b}}+\varkappa_{\mathbf{b}}^{\mathbf{a}}d\theta^{\mathbf{b}%
}+\pounds _{\xi}\omega_{\mathbf{b}}^{\mathbf{a}}\wedge\theta^{\mathbf{b}%
}+\varkappa_{\mathbf{c}}^{\mathbf{b}}\omega_{\mathbf{b}}^{\mathbf{a}}%
\wedge\theta^{\mathbf{b}}. \label{7.46}%
\end{align}
Also, using Eq.(\ref{7.45}) we have
\begin{equation}
\pounds _{\xi}\Theta^{\mathbf{a}}=\varkappa_{\mathbf{b}}^{\mathbf{a}}%
d\theta^{\mathbf{b}}+\varkappa_{\mathbf{c}}^{\mathbf{b}}\omega_{\mathbf{b}%
}^{\mathbf{a}}\wedge\theta^{\mathbf{c}}. \label{7.47}%
\end{equation}
From Eqs.(\ref{7.46}) and (\ref{7.47} it follows that $\pounds _{\xi}%
\omega_{\mathbf{b}}^{\mathbf{a}}\wedge\theta^{\mathbf{b}}=\varkappa
_{\mathbf{c}}^{\mathbf{a}}\omega_{\mathbf{b}}^{\mathbf{c}}\wedge
\theta^{\mathbf{b}}-\varkappa_{\mathbf{c}}^{\mathbf{b}}\omega_{\mathbf{b}%
}^{\mathbf{a}}\wedge\theta^{\mathbf{b}}-d\left(  \varkappa_{\mathbf{b}%
}^{\mathbf{a}}\right)  \wedge\theta^{\mathbf{b}}$, or
\begin{equation}
\pounds _{\xi}\omega_{\mathbf{b}}^{\mathbf{a}}=\varkappa_{\mathbf{c}%
}^{\mathbf{a}}\omega_{\mathbf{b}}^{\mathbf{c}}-\varkappa_{\mathbf{b}%
}^{\mathbf{c}}\omega_{\mathbf{c}}^{\mathbf{a}}-d\varkappa_{\mathbf{b}%
}^{\mathbf{a}} \label{7.48}%
\end{equation}

Thus, recalling Eq.(\ref{7.30}) we finally have that $\pounds _{\xi}%
\omega_{\mathbf{b}}^{\mathbf{a}}=%
\mbox{\boldmath{$\delta$}}%
_{v}\omega_{\mathbf{b}}^{\mathbf{a}}$.
\end{proof}

\begin{corollary}
\label{for cons}For any $x\in M$, $\theta_{\mathbf{b}}\cdot\mathbf{L}_{\xi
}\theta^{\mathbf{a}}$ is an element of $\mathrm{spin}_{1,3}^{e}$, if and only
if, $\pounds _{\xi}\mathbf{g}=0$.
\end{corollary}

\begin{proof}
The Lie covariant derivative of $\theta^{\mathbf{a}}$ is given by
\begin{align}
\mathbf{L}_{\xi}\theta^{\mathbf{a}}  &  =\xi^{\ast}\mathbin\lrcorner
\mathbf{D}\theta^{\mathbf{a}}+\mathbf{D}\left(  \mathbf{\xi}^{\ast}\cdot
\theta^{\mathbf{a}}\right) \nonumber\\
&  =\xi^{\ast}\mathbin\lrcorner\left(  d\theta^{\mathbf{a}}+\omega
_{\mathbf{b}}^{\mathbf{a}}\wedge\theta^{\mathbf{b}}\right)  +d\left(
\mathbf{\xi}^{\ast}\cdot\theta^{\mathbf{a}}\right)  +\omega_{\mathbf{b}%
}^{\mathbf{a}}\left(  \mathbf{\xi}^{\ast}\cdot\theta^{\mathbf{b}}\right)
\nonumber\\
&  =\pounds _{\xi}\theta^{\mathbf{a}}+(\xi^{\ast}\cdot\omega_{\mathbf{b}%
}^{\mathbf{a}})\theta^{\mathbf{b}}-(\xi^{\ast}\cdot\theta^{\mathbf{b}}%
)\omega_{\mathbf{b}}^{\mathbf{a}}+\omega_{\mathbf{b}}^{\mathbf{a}}\left(
\xi^{\ast}\cdot\theta^{\mathbf{b}}\right) \nonumber\\
&  =\pounds _{\xi}\theta^{\mathbf{a}}+(\xi^{\ast}\cdot\omega_{\mathbf{b}%
}^{\mathbf{a}})\theta^{\mathbf{b}}\nonumber\\
&  =\left(  \varkappa_{\mathbf{b}}^{\mathbf{a}}+\xi^{\ast}\cdot\omega
_{\mathbf{b}}^{\mathbf{a}}\right)  \theta^{\mathbf{b}}, \label{7.49}%
\end{align}
where we put $\pounds _{\xi}\theta^{\mathbf{a}}=\varkappa_{\mathbf{b}%
}^{\mathbf{a}}\theta^{\mathbf{b}}$. Then,%
\begin{equation}
\theta_{\mathbf{b}}\cdot\mathbf{L}_{\xi}\theta^{\mathbf{a}}=\varkappa
_{\mathbf{b}}^{\mathbf{a}}+\xi^{\ast}\cdot\omega_{\mathbf{b}}^{\mathbf{a}}.
\label{7.50}%
\end{equation}
Now, we have already shown above that for any $x\in M$, the matrix of the
$\xi^{\ast}\cdot\omega_{\mathbf{b}}^{\mathbf{a}}$ is an element of
$\mathrm{spin}_{1,3}^{e}$ and then, $\theta_{\mathbf{b}}\cdot\mathbf{L}_{\xi
}\theta^{\mathbf{a}}$ will be an element of $\mathrm{spin}_{1,3}^{e}$ if and
only if the matrix of the $\varkappa_{\mathbf{b}}^{\mathbf{a}}$ is an element
of $\mathrm{spin}_{1,3}^{e}$. The corollary is proved.
\end{proof}

\begin{lemma}
\label{step2}If \ $\pounds _{\xi}\mathbf{g}=0$ and $\pounds _{\xi}\Theta=0$
then we have the identity
\begin{equation}
\mathbf{D}\left(  \theta_{\mathbf{b}}\cdot\mathbf{L}_{\xi}\theta^{\mathbf{a}%
}\right)  +\xi^{\ast}\mathbin\lrcorner\mathcal{R}_{\mathbf{b}}^{\mathbf{a}}=0.
\label{7.51}%
\end{equation}

\end{lemma}

\begin{proof}
Using the definitions of the exterior covariant derivative and the Lie
covariant derivative we have%
\begin{align*}
\mathbf{D}\left(  \theta_{\mathbf{b}}\cdot\mathbf{L}_{\xi}\theta^{\mathbf{a}%
}\right)   &  =d\left(  \theta_{\mathbf{b}}\cdot\mathbf{L}_{\xi}%
\theta^{\mathbf{a}}\right)  +\omega_{\mathbf{b}}^{\mathbf{c}}(\theta
_{\mathbf{c}}\cdot\mathbf{L}_{\xi}\theta^{\mathbf{a}})-\omega_{\mathbf{c}%
}^{\mathbf{a}}(\theta_{\mathbf{b}}\cdot\mathbf{L}_{\xi}\theta^{\mathbf{c}})\\
&  =d\left\{  \theta_{\mathbf{b}}\cdot\left[  \mathbf{\pounds }_{\xi}%
\theta^{\mathbf{a}}+(\xi^{\ast}\cdot\omega_{\mathbf{c}}^{\mathbf{a}}%
)\theta^{\mathbf{c}}\right]  \right\} \\
&  +\left\{  \theta_{\mathbf{d}}\cdot\left[  \mathbf{\pounds }_{\xi}%
\theta^{\mathbf{a}}+(\xi^{\ast}\cdot\omega_{\mathbf{c}}^{\mathbf{a}}%
)\theta^{\mathbf{c}}\right]  \right\}  \omega_{\mathbf{b}}^{\mathbf{d}}\\
&  -\left\{  \theta_{\mathbf{b}}\cdot\left[  \mathbf{\pounds }_{\xi}%
\theta^{\mathbf{d}}+(\xi^{\ast}\cdot\omega_{\mathbf{c}}^{\mathbf{d}}%
)\theta^{\mathbf{c}}\right]  \right\}  \omega_{\mathbf{d}}^{\mathbf{a}},
\end{align*}
i.e.,
\begin{align}
\mathbf{D}\left(  \theta_{\mathbf{b}}\cdot\mathbf{L}_{\xi}\theta^{\mathbf{a}%
}\right)   &  =\pounds _{\xi}\omega_{\mathbf{b}}^{\mathbf{a}}-\xi^{\ast
}\mathbin\lrcorner\left(  d\omega_{\mathbf{b}}^{\mathbf{a}}+\omega
_{\mathbf{c}}^{\mathbf{a}}\wedge\omega_{\mathbf{b}}^{\mathbf{c}}\right)
\label{7.52}\\
&  +d(\theta_{\mathbf{b}}\cdot\mathbf{\pounds }_{\xi}\theta^{\mathbf{a}%
})+\omega_{\mathbf{b}}^{\mathbf{c}}(\theta_{\mathbf{c}}\cdot\mathbf{\pounds }%
_{\xi}\theta^{\mathbf{a}})-(\theta_{\mathbf{b}}\cdot\mathbf{\pounds }_{\xi
}\theta^{\mathbf{c}})\omega_{\mathbf{c}}^{\mathbf{a}}.\nonumber
\end{align}

If, $\pounds _{\xi}\mathbf{g}=0$, then for any $x\in M$, $\theta_{\mathbf{b}%
}\cdot\mathbf{\pounds }_{\xi}\theta^{\mathbf{a}}\in\mathrm{spin}_{1,3}^{e}$
and the second line of Eq.(\ref{7.52}) is an infinitesimal Lorentz
transformation of the $\omega_{\mathbf{b}}^{\mathbf{a}}$. If besides that,
also $\pounds _{\xi}\Theta=0$ then $\pounds _{\xi}\omega_{\mathbf{b}%
}^{\mathbf{a}}=%
\mbox{\boldmath{$\delta$}}%
_{v}\omega_{\mathbf{b}}^{\mathbf{a}}$ and then the first term on the second
member of Eq.(\ref{7.52}) cancels the term in the second line. Then, taking
into account Cartan's second structure equation the proposition is
proved.\medskip
\end{proof}

\begin{proof}
(\textbf{Proposition \ref{killing propos}}). We are now in conditions of
presenting a proof of the Proposition \ref{killing propos}. In order to do
that we combine the results of \ Lemmas \ref{step1} and \ref{step2} \ with the
identities given by Eqs.(\ref{7.37}) and (\ref{7.39}). We get,%

\begin{align*}
d[(\xi^{\ast}\cdot\theta^{\mathbf{a}})\star\mathcal{T}_{\mathbf{a}}]  &
=\mathbf{D}[(\xi^{\ast}\cdot\theta^{\mathbf{a}})\star\mathcal{T}_{\mathbf{a}%
}]\\
&  =\mathbf{D}(\xi^{\ast}\cdot\theta^{\mathbf{a}})\wedge\star\mathcal{T}%
_{\mathbf{a}}+(\xi^{\ast}\cdot\theta^{\mathbf{a}})\mathbf{D}\star
\mathcal{T}_{\mathbf{a}}\\
&  =\mathbf{L}_{\xi}\theta^{\mathbf{a}}\wedge\star\mathcal{T}_{\mathbf{a}%
}-\left(  \xi^{\ast}\lrcorner\Theta^{\mathbf{a}}\right)  \wedge\star
\mathcal{T}_{\mathbf{a}}+(\xi^{\ast}\cdot\theta^{\mathbf{a}})\mathbf{D}%
\star\mathcal{T}_{\mathbf{a}},
\end{align*}
i.e.,%
\begin{equation}
d[(\xi^{\ast}\cdot\theta^{\mathbf{a}})\star\mathcal{T}_{\mathbf{a}%
}]=\mathbf{L}_{\xi}\theta^{\mathbf{a}}\wedge\star\mathcal{T}_{\mathbf{a}%
}-\star\mathbf{J}_{\mathbf{b}}^{\mathbf{a}}\wedge\left(  \xi^{\ast}%
\lrcorner\mathcal{R}_{\mathbf{a}}^{\mathbf{b}}\right)  . \label{7.53}%
\end{equation}

Observe now that if $A\in\sec\bigwedge\nolimits^{1}TM\hookrightarrow
\sec\mathcal{C}\ell(M,\mathtt{g})$ then, $\theta^{\mathbf{a}}\wedge
(\theta_{\mathbf{a}}\cdot A)=A$. This permits us to write Eq.(\ref{7.53}) as%
\begin{equation}
d[(\xi^{\ast}\cdot\theta^{\mathbf{a}})\star\mathcal{T}_{\mathbf{a}}]=-\left(
\theta_{\mathbf{b}}\cdot\mathbf{L}_{\xi}\theta^{\mathbf{a}}\right)
\wedge\star\mathcal{T}_{\mathbf{a}}\wedge\theta^{\mathbf{b}}-\star
\mathbf{J}_{\mathbf{b}}^{\mathbf{a}}\wedge\left(  \xi^{\ast}\lrcorner
\mathcal{R}_{\mathbf{a}}^{\mathbf{b}}\right)  . \label{7.54}%
\end{equation}

If $\pounds _{\xi}\mathtt{\mathbf{g}}=0$, we have by the Corollary of
Proposition \ref{step1} that for any $x\in M$, $\theta_{\mathbf{b}}%
\cdot\mathbf{L}_{\xi}\theta^{\mathbf{a}}\in\mathrm{spin}_{1,3}^{e}$. In that
case, we can write Eq.(\ref{7.54}) as%
\begin{align}
d[(\xi^{\ast}\cdot\theta^{\mathbf{a}})\star\mathcal{T}_{\mathbf{a}}]  &
=-\frac{1}{2}\left(  \theta_{\mathbf{b}}\cdot\mathbf{L}_{\xi}\theta
^{\mathbf{a}}\right)  \wedge\lbrack\star\mathcal{T}_{\mathbf{a}}\wedge
\theta^{\mathbf{b}}-\star\mathcal{T}^{\mathbf{b}}\wedge\theta_{\mathbf{a}%
}]-\star\mathbf{J}_{\mathbf{b}}^{\mathbf{a}}\wedge\left(  \xi^{\ast}%
\lrcorner\mathcal{R}_{\mathbf{a}}^{\mathbf{b}}\right) \nonumber\\
&  =-\left(  \theta_{\mathbf{b}}\cdot\mathbf{L}_{\xi}\theta^{\mathbf{a}%
}\right)  \wedge\mathbf{D\star J}_{\mathbf{a}}^{\mathbf{b}}-\star
\mathbf{J}_{\mathbf{b}}^{\mathbf{a}}\wedge\left(  \xi^{\ast}\lrcorner
\mathcal{R}_{\mathbf{a}}^{\mathbf{b}}\right)  . \label{7.55}%
\end{align}

On the other hand, if $\pounds _{\xi}\Theta=0$, in view of Proposition
\ref{step2} we can write%
\begin{align}
d[(\xi^{\ast}\cdot\theta^{\mathbf{a}})\star\mathcal{T}_{\mathbf{a}}]  &
=-\mathbf{D}\left(  \theta_{\mathbf{b}}\cdot\mathbf{L}_{\xi}\theta
^{\mathbf{a}}\right)  \wedge\star J_{\mathbf{a}}^{\mathbf{b}}-\left(
\theta_{\mathbf{b}}\cdot\mathbf{L}_{\xi}\theta^{\mathbf{a}}\right)
\wedge\mathbf{D\star J}_{\mathbf{a}}^{\mathbf{b}}\nonumber\\
&  =-\mathbf{D[}\left(  \theta_{\mathbf{b}}\cdot\mathbf{L}_{\xi}%
\theta^{\mathbf{a}}\right)  \wedge\star J_{\mathbf{a}}^{\mathbf{b}%
}]=-d\mathbf{[}\left(  \theta_{\mathbf{b}}\cdot\mathbf{L}_{\xi}\theta
^{\mathbf{a}}\right)  \wedge\star J_{\mathbf{a}}^{\mathbf{b}}]. \label{7.56}%
\end{align}

Finally, if $\pounds _{\xi}\mathtt{\mathbf{g}}=0$ and $\pounds _{\xi}\Theta=0$
we have
\[
d[(\xi^{\ast}\cdot\theta^{\mathbf{a}})\star\mathcal{T}_{\mathbf{a}}+\left(
\theta_{\mathbf{b}}\cdot\mathbf{L}_{\xi}\theta^{\mathbf{a}}\right)
\wedge\star J_{\mathbf{a}}^{\mathbf{b}}]=0,
\]
which is the result we wanted to prove.
\end{proof}

\section{Pseudo Potentials in General Relativity}

As we already said, in Einstein's gravitational theory (General Relativity)
each gravitational field is modelled by a Lorentzian spacetime $\mathfrak{M}%
=(M,\mathbf{g},D,\tau_{\mathbf{g}},\uparrow)$. The `gravitational field'
$\mathtt{\mathbf{g}}$ is determined through Einstein's equations by the
energy-momentum of the matter fields $\phi^{A}$, $A=1,2,...,m$, living in
$\mathfrak{M}$. As shown in details in, e.g., \cite{roqui2005,rodoliv2006}
Einstein's equations can be written using the Clifford bundle formalism in
terms of the fields $\theta^{\mathbf{a}}\in\sec\bigwedge\nolimits^{1}T^{\ast
}M\hookrightarrow\sec\mathcal{C}\ell(M,\mathtt{g})$, where $\{\theta
^{\mathbf{a}}\}$ is an \textit{orthonormal }basis of $T^{\ast}M$ as%
\begin{equation}
-({\mbox{\boldmath$\partial$}}\cdot{\mbox{\boldmath$\partial$}})\theta
^{\mathbf{a}}+{\mbox{\boldmath$\partial$}}\wedge({\mbox{\boldmath$\partial$}}%
\cdot\theta^{\mathbf{a}})+{\mbox{\boldmath$\partial$}}\lrcorner
({\mbox{\boldmath$\partial$}}\wedge\theta^{\mathbf{a}})+\frac{1}{2}%
\mathcal{T}\theta^{\mathbf{a}}=\mathcal{T}^{\mathbf{a}}, \label{7.56a}%
\end{equation}
where ${\mbox{\boldmath$\partial$}}=\theta^{\mathbf{a}}D_{\mathbf{e}%
_{\mathbf{a}}}$ is the Dirac operator acting on sections of the Clifford
bundle. An explicit Lagrangian giving that equation\footnote{Eq.(\ref{7.58})
is equivalent to a Lagragian density first introduced by \cite{moller3}.},
which differs from the original Einstein-Hilbert Lagrangian\footnote{Recall
that the Einstein-Hilbert Lagrangian density is $\mathcal{L}_{EH}=\frac{1}%
{2}R\tau_{\mathtt{g}}=\frac{1}{2}\mathcal{R}_{\mathbf{cd}}\wedge\star
(\theta^{\mathbf{c}}\wedge\theta^{\mathbf{d}})$ and $\mathcal{L}%
_{EH}=\mathcal{L}_{g}-d(\theta^{\mathbf{a}}\wedge\star d\theta_{\mathbf{a}}%
)$.} by an exact differential is%

\begin{equation}
\mathcal{L}_{g}=-\frac{1}{2}d\theta^{\mathbf{a}}\wedge\star d\theta
_{\mathbf{a}}+\frac{1}{2}\delta\theta^{\mathbf{a}}\wedge\star\delta
\theta_{\mathbf{a}}+\frac{1}{4}\left(  d\theta^{\mathbf{a}}\wedge
\theta_{\mathbf{a}}\right)  \wedge\star\left(  d\theta^{\mathbf{b}}%
\wedge\theta_{\mathbf{b}}\right)  . \label{7.58}%
\end{equation}

The total Lagrangian density of the gravitational field and the matter fields
can\ then be written as
\begin{equation}
\mathcal{L}=\mathcal{L}_{g}+\mathcal{L}_{m}, \label{7.57}%
\end{equation}
where $\mathcal{L}_{m}(\theta^{\mathbf{a}},d\theta^{\mathbf{a}},\phi^{A}%
,d\phi^{A})$ is the matter Lagrangian.

Now, variation of $\mathcal{L}$ with respect to the the fields $\theta
^{\mathbf{a}}$ yields after a very long calculation (see, e.g.,
\cite{roqui2005}) the following Euler-Lagrange equations%
\begin{equation}
-\star\mathcal{G}^{\mathbf{a}}=\frac{\partial\mathcal{L}_{g}}{\partial
\theta_{a}}+d\left(  \frac{\partial\mathcal{L}_{g}}{\partial d\theta_{a}%
}\right)  =\star t^{\mathbf{a}}+d\star\mathcal{S}^{\mathbf{a}}=-\star
\mathcal{T}^{\mathbf{a}}, \label{7.59}%
\end{equation}
where $\mathcal{G}^{\mathbf{a}}=(\mathcal{R}^{\mathbf{a}}-\frac{1}{2}%
R\theta^{\mathbf{a}})\in\sec\bigwedge\nolimits^{1}T^{\ast}M\hookrightarrow
\mathcal{C\ell}\left(  T^{\ast}M,\mathtt{g}\right)  $ are the Einstein
$1$-forms, $\mathcal{R}^{\mathbf{a}}=R_{\mathbf{b}}^{\mathbf{a}}%
\theta^{\mathbf{b}}$ $\in\sec\bigwedge\nolimits^{1}T^{\ast}M\hookrightarrow
\mathcal{C\ell}\left(  T^{\ast}M,\mathtt{g}\right)  $ are the Ricci $1$-forms,
$R$ is the scalar curvature, $\star\mathcal{T}^{\mathbf{a}}=\frac
{\partial\mathcal{L}_{m}}{\partial\theta_{\mathbf{a}}}\in\sec\bigwedge
\nolimits^{1}T^{\ast}M$ $\hookrightarrow\mathcal{C\ell}\left(  T^{\ast
}M,\mathtt{\mathbf{g}}\right)  $ are the energy-momentum $1$-forms of the
matter fields\footnote{Recall that due to our conventions in the writing of
Einstein equations the true physical energy-momentum densities are $\star
T^{\mathbf{a}}=-\star\mathcal{T}^{\mathbf{a}}$. The objects $\star
t^{\mathbf{a}}$ and $d\star S^{\mathbf{a}}$ are more easily found by variation
of $\mathcal{L}_{EH}$ instead of the variation of $\mathcal{L}_{g}$, which of
course, give the same equations of motion.\smallskip}, and where
\begin{align}
\star\mathcal{S}^{\mathbf{c}}  &  =\frac{\partial\mathcal{L}_{m}}{\partial
d\theta_{\mathbf{a}}}=\frac{1}{2}%
\mbox{\boldmath{$\omega$}}%
_{\mathbf{ab}}\wedge\star(\theta^{\mathbf{a}}\wedge\theta^{\mathbf{b}}%
\wedge\theta^{\mathbf{c}})\in\sec\bigwedge\nolimits^{2}T^{\ast}%
M\hookrightarrow\mathcal{C\ell}\left(  T^{\ast}M,\mathtt{\mathbf{g}}\right)
,\nonumber\\
\star t_{\mathbf{\ }}^{\mathbf{c}}  &  =\frac{\partial\mathcal{L}_{m}%
}{\partial\theta_{\mathbf{a}}}=-\frac{1}{2}%
\mbox{\boldmath{$\omega$}}%
_{\mathbf{ab}}\wedge\lbrack%
\mbox{\boldmath{$\omega$}}%
_{\mathbf{d}}^{\mathbf{c}}\wedge\star(\theta^{\mathbf{a}}\wedge\theta
^{\mathbf{b}}\wedge\theta^{\mathbf{d}})+%
\mbox{\boldmath{$\omega$}}%
_{\mathbf{d}}^{\mathbf{b}}\wedge\star(\theta^{\mathbf{a}}\wedge\theta
^{\mathbf{d}}\wedge\theta^{\mathbf{c}})]\label{7.10.17}\\
&  \in\sec\bigwedge\nolimits^{3}T^{\ast}M\hookrightarrow\mathcal{C\ell}\left(
T^{\ast}M,\mathtt{\mathbf{g}}\right)  .\nonumber
\end{align}

The proof that the second and third members of Eq.(\ref{7.59}) are equal
follows at once from the fact that the connection $1$-forms of the Levi-Civita
connection of $\mathbf{g}$ can be written as it is trivial to verify as%
\begin{equation}
\omega^{\mathbf{cd}}=\frac{1}{2}\left[  \theta^{\mathbf{d}}\lrcorner
d\theta^{\mathbf{c}}-\theta^{\mathbf{c}}\lrcorner d\theta^{\mathbf{d}}%
+\theta^{\mathbf{c}}\lrcorner\left(  \theta^{\mathbf{d}}\lrcorner
d\theta_{\mathbf{a}}\right)  \theta^{\mathbf{a}}\right]  , \label{w equation}%
\end{equation}
and that
\begin{equation}
\star\mathcal{G}^{\mathbf{d}}=-\frac{1}{2}\mathcal{R}_{\mathbf{ab}}\wedge
\star(\theta^{\mathbf{a}}\wedge\theta^{\mathbf{b}}\wedge\theta^{\mathbf{d}}).
\label{7.10.18}%
\end{equation}

Indeed, we can write%
\begin{align}
\frac{1}{2}\mathcal{R}_{\mathbf{ab}}\wedge\star(\theta^{\mathbf{a}}%
\wedge\theta^{\mathbf{b}}\wedge\theta^{\mathbf{d}})  &  =-\frac{1}{2}%
\star\lbrack\mathcal{R}_{\mathbf{ab}}\lrcorner(\theta^{\mathbf{a}}\wedge
\theta^{\mathbf{b}}\wedge\theta^{\mathbf{d}})]\nonumber\\
&  =-\frac{1}{2}R_{\mathbf{abcd}}\star\lbrack(\theta^{\mathbf{c}}\wedge
\theta^{\mathbf{d}})\lrcorner(\theta^{\mathbf{a}}\wedge\theta^{\mathbf{b}%
}\wedge\theta^{\mathbf{d}})]\nonumber\\
&  =-\star(\mathcal{R}^{\mathbf{d}}-\frac{1}{2}R\theta^{\mathbf{d}}).
\label{7.10.19}%
\end{align}

On the other hand we have,%
\begin{align}
-  &  2\star\mathcal{G}^{\mathbf{d}}=d%
\mbox{\boldmath{$\omega$}}%
_{\mathbf{ab}}\wedge\star(\theta^{\mathbf{a}}\wedge\theta^{\mathbf{b}}%
\wedge\theta^{\mathbf{d}})+%
\mbox{\boldmath{$\omega$}}%
_{\mathbf{ac}}\wedge%
\mbox{\boldmath{$\omega$}}%
_{\mathbf{b}}^{\mathbf{c}}\wedge\star(\theta^{\mathbf{a}}\wedge\theta
^{\mathbf{b}}\wedge\theta^{\mathbf{d}})\nonumber\\
&  =d[%
\mbox{\boldmath{$\omega$}}%
_{\mathbf{ab}}\wedge\star(\theta^{\mathbf{a}}\wedge\theta^{\mathbf{b}}%
\wedge\theta^{\mathbf{d}})]+%
\mbox{\boldmath{$\omega$}}%
_{\mathbf{ab}}\wedge d\star(\theta^{\mathbf{a}}\wedge\theta^{\mathbf{b}}%
\wedge\theta^{\mathbf{d}})\nonumber\\
&  +%
\mbox{\boldmath{$\omega$}}%
_{\mathbf{ac}}\wedge%
\mbox{\boldmath{$\omega$}}%
_{\mathbf{b}}^{\mathbf{c}}\wedge\star(\theta^{\mathbf{a}}\wedge\theta
^{\mathbf{b}}\wedge\theta^{\mathbf{d}})\nonumber\\
&  =d[%
\mbox{\boldmath{$\omega$}}%
_{\mathbf{ab}}\wedge\star(\theta^{\mathbf{a}}\wedge\theta^{\mathbf{b}}%
\wedge\theta^{\mathbf{d}})]-%
\mbox{\boldmath{$\omega$}}%
_{\mathbf{ab}}\wedge%
\mbox{\boldmath{$\omega$}}%
_{\mathbf{p}}^{\mathbf{a}}\wedge\star(\theta^{\mathbf{p}}\wedge\theta
^{\mathbf{b}}\wedge\theta^{\mathbf{d}})\nonumber\\
&  -%
\mbox{\boldmath{$\omega$}}%
_{\mathbf{ab}}\wedge%
\mbox{\boldmath{$\omega$}}%
_{\mathbf{p}}^{\mathbf{b}}\star(\theta^{\mathbf{a}}\wedge\theta^{\mathbf{p}%
}\wedge\theta^{\mathbf{d}})-%
\mbox{\boldmath{$\omega$}}%
_{\mathbf{ab}}\wedge%
\mbox{\boldmath{$\omega$}}%
_{\mathbf{p}}^{\mathbf{d}}\wedge\star(\theta^{\mathbf{a}}\wedge\theta
^{\mathbf{b}}\wedge\theta^{\mathbf{p}})]\nonumber\\
&  +%
\mbox{\boldmath{$\omega$}}%
_{\mathbf{ac}}\wedge%
\mbox{\boldmath{$\omega$}}%
_{\mathbf{b}}^{\mathbf{c}}\wedge\star(\theta^{\mathbf{a}}\wedge\theta
^{\mathbf{b}}\wedge\theta^{\mathbf{d}})\nonumber\\
&  =d[%
\mbox{\boldmath{$\omega$}}%
_{\mathbf{ab}}\wedge\star(\theta^{\mathbf{a}}\wedge\theta^{\mathbf{b}}%
\wedge\theta^{\mathbf{d}})]-%
\mbox{\boldmath{$\omega$}}%
_{\mathbf{ab}}\wedge\lbrack%
\mbox{\boldmath{$\omega$}}%
_{\mathbf{p}}^{\mathbf{d}}\wedge\star(\theta^{\mathbf{a}}\wedge\theta
^{\mathbf{b}}\wedge\theta^{\mathbf{p}})\nonumber\\
&  +%
\mbox{\boldmath{$\omega$}}%
_{\mathbf{p}}^{\mathbf{b}}\wedge\star(\theta^{\mathbf{a}}\wedge\theta
^{\mathbf{p}}\wedge\theta^{\mathbf{d}})]\nonumber\\
&  =2(d\mathop\star\mathcal{S}^{\mathbf{d}}+\star t^{\mathbf{d}}).
\label{7.10.20}%
\end{align}

Now, we can write Einstein's equation in a very interesting, but
\textit{dangerous} form\footnote{Eq.(\ref{7.10.21}) is known in recent
literature of GR as Sparling equations \ref{szabados} because it appears (in
an equivalent form) in a preprint \ref{sparling} of 1982 by that author.
However, it already appeared early, e.g., in a 1978 paper by Thirring and
Wallner \cite{thiwal}.}, i.e.:%
\begin{equation}
-d\star\mathcal{S}^{\mathbf{a}}=\star\mathcal{T}^{\mathbf{a}}+\star
t^{\mathbf{a}}. \label{7.10.21}%
\end{equation}

In writing Einstein's equations in that way, we have associated to the
gravitational field a set of $2$-form fields $\star\mathcal{S}^{\mathbf{a}}$
called \textit{superpotentials} that have as sources the currents
$(\star\mathcal{T}^{\mathbf{a}}+\star t^{\mathbf{a}})$. However,
superpotentials are not uniquely defined since, e.g., superpotentials
\ $(\star\mathcal{S}^{\mathbf{a}}+\star\alpha^{\mathbf{a}})$, with
$\star\alpha^{\mathbf{a}}$ closed, i.e., $d\star\alpha^{\mathbf{a}}=0$ give
the same second member for Eq.(\ref{7.10.21}).

\subsection{Is There Any Energy-Momentum Conservation Law in GRT?}

Why did we say that Eq.(\ref{7.10.21}) is a dangerous one?

The reason is that if we are ignorant of the discussion of the previous
section we may be led to think that we have discovered a conservation law for
the energy momentum of matter plus gravitational field, since from
Eq.(\ref{7.10.21}) it follows that
\begin{equation}
d(\star\mathcal{T}^{\mathbf{a}}+\star t^{\mathbf{a}})=0. \label{7.10.22}%
\end{equation}
This thought however is only an example of wishful thinking, because the
$\star t^{\mathbf{a}}$ \ depends on the connection (see Eq.(\ref{7.10.17}))
and thus are gauge dependent. They do not have the same tensor transformation
law as the $\star\mathcal{T}^{\mathbf{a}}$. So, Stokes theorem cannot be used
to derive from Eq.(\ref{7.10.22}) conserved quantities that are independent of
the gauge, which is clear. However---and this is less known---Stokes theorem
also cannot be used to derive conclusions that are independent of the local
coordinate \textit{chart} used to perform calculations \cite{boro}. In fact,
the currents $\star t^{\mathbf{a}}$ are nothing more than the old pseudo
energy-momentum tensor of Einstein in a new dress. Non recognition of this
fact can lead to many misunderstandings. We present some of them in what
follows, in order to call our readers' attention of potential errors of
inference that can be done when we use sophisticated mathematical formalisms
without a perfect domain of their contents.

\textbf{(i)}\ First, it is easy to see that from Eq.(\ref{7.59}) it follows
that \cite{mtw}
\begin{equation}
\mathbf{D\star}\mathfrak{G}=\mathbf{D}\star\mathfrak{T}=0, \label{7.10.22.0}%
\end{equation}
where $\mathbf{\star}\mathfrak{G}=\mathbf{e}_{\mathbf{a}}\otimes
\star\mathcal{G}^{\mathbf{a}}$ $\in\sec TM\otimes\sec\bigwedge\nolimits^{3}%
T^{\ast}M$ and $\star\mathfrak{T}=\mathbf{e}_{\mathbf{a}}\otimes
\star\mathcal{T}^{\mathbf{a}}\in\sec TM\otimes\sec\bigwedge\nolimits^{3}%
T^{\ast}M$ and where%
\begin{equation}
\mathbf{D\star}\mathfrak{G}:=\mathbf{e}_{\mathbf{a}}\otimes\mathbf{D}%
\star\mathcal{G}^{\mathbf{a}}\text{, }\mathbf{D}\star\mathfrak{T}%
=\mathbf{e}_{\mathbf{a}}\otimes\mathbf{D}\star\mathcal{T}^{\mathbf{a}}
\label{7.10.22n}%
\end{equation}
and $\mathbf{D}$ is the exterior covariant derivative of index valued forms
(\cite{benntucker,rodoliv2006}). Now, in \cite{mtw} it is written (without
proof) a \ `Stokes theorem' \medskip%
\begin{equation}%
\begin{tabular}
[c]{|c|}\hline
$%
{\displaystyle\int\limits_{{\footnotesize 4}\text{-cube}}}
\mathbf{D\star}\mathfrak{T}\mathbf{=}%
{\displaystyle\int\limits_{\substack{{\footnotesize 3-}\text{ boundary}%
\\\text{ of this }{\footnotesize 4}\text{-cube}}}}
\star\mathfrak{T.}$\\\hline
\end{tabular}
\ \label{mtw}%
\end{equation}

Not a single proof (which we can consider as valid) of Eq.(\ref{mtw}) which
appears also in many other texts and scientific papers as, e.g., in
\cite{dalton,vatorr1} has been given in any paper we know. The reason is the
following. The first member of Eq.(\ref{mtw}) is no more than
\begin{equation}%
{\displaystyle\int\limits_{{\footnotesize 4}\text{-cube}}}
\mathbf{e}_{\mathbf{a}}\otimes(d\star\mathcal{T}^{\mathbf{a}}+\omega
_{\mathbf{b}}^{\mathbf{a}}\wedge\star\mathcal{T}^{\mathbf{b}}).
\label{7.10.22.01}%
\end{equation}
Thus it is necessary to explain what is the meaning (if any) of the integral.
Since the integrand is a sum of tensor fields, this integral says that we are
\textit{adding} tensors belonging to the tensor spaces of different spacetime
points. As it is well known, this cannot be done in general, unless there is a
way of identifying the tensor spaces at different spacetime points. This
requires, of course, the introduction of additional structure on the spacetime
representing a given gravitational field, and such extra structure is lacking
in Einstein theory. We must conclude that Eq.(\ref{mtw}) do not express any
conservation law, for it lacks as yet, a precise mathematical meaning.

In Einstein theory possible superpotentials are, of course, the $\star
\mathcal{S}^{\mathbf{a}}$ that we identified above (Eq.(\ref{7.10.17})), with
\begin{equation}
\star\mathcal{S}_{\mathbf{c}}=[\frac{1}{2}\omega_{\mathbf{ab}}\lrcorner
(\theta^{\mathbf{a}}\wedge\theta^{\mathbf{b}}\wedge\theta_{\mathbf{c}}%
)]\theta^{\mathbf{5}}. \label{7.10.22.1.1}%
\end{equation}

Then, if we integrate Eq.(\ref{7.10.21}) over a \ `certain finite
$3$-dimensional volume', say a ball $B$, and use Stokes theorem we
have\footnote{The reason for the factor $8\pi$ in Eq.(\ref{7.10.22.2}) is that
we choose units where the numerical value gravitational constant $8\pi
G/c^{4}$ is $1$, where $G$ is Newton gravitational constant.}%
\begin{equation}
P^{\mathbf{a}}:=-\frac{1}{8\pi}%
{\displaystyle\int\limits_{B}}
\star\left(  \mathcal{T}^{\mathbf{a}}+t^{\mathbf{a}}\right)  =\frac{1}{8\pi}%
{\displaystyle\int\limits_{B}}
\star\left(  T^{\mathbf{a}}-t^{\mathbf{a}}\right)  =\frac{1}{8\pi}%
{\displaystyle\int\limits_{\partial B}}
\star\mathcal{S}^{\mathbf{a}}. \label{7.10.22.2}%
\end{equation}

In particular the energy or (\textit{inertial mass}) of the gravitational
field plus matter generating the field is defined by
\begin{equation}
P^{\mathbf{0}}=E=m_{\mathbf{I}}=\frac{1}{8\pi}\lim_{R\rightarrow\infty}%
{\displaystyle\int\limits_{\partial B}}
\star\mathcal{S}^{\mathbf{0}}. \label{7.10.22.3'}%
\end{equation}

\textbf{(ii) }Now, a frequent misunderstanding is the following. Suppose that
in a \textit{given }hypothetical gravitational theory\ there exists an
energy-momentum conservation law for matter plus the gravitational field
expressed in the form of Eq.(\ref{7.10.22}), where $\mathcal{T}^{\mathbf{a}}$
are the energy-momentum 1-forms of matter and $t^{\mathbf{a}}$ are
\textit{true}\footnote{This means that the $t^{\mathbf{a}}$ in the
hypothetical theory are not i pseudo 1-forms, as is the case in Einstein's
theory.} energy-momentum 1-forms of the gravitational field. This means that
the $3$-forms $(\star\mathcal{T}^{\mathbf{a}}+\star t^{\mathbf{a}})$ are
closed, i.e., they satisfy Eq.(\ref{7.10.22}). Is this enough to warrant that
the energy of a closed universe is zero? Well, that would be the case if
starting from Eq.(\ref{7.10.22}) we could jump to an equation like
Eq.(\ref{7.10.21}) and then to Eq.(\ref{7.10.22.3'}) (as done, e.g., in
\cite{thiwal}). But that sequence of inferences in general cannot be done, for
indeed, as it is well known, it is not the case that closed three forms are
always exact. Take, for example, a closed universe with topology
$\mathbb{R\times}S^{3}$. In this case $B=S^{3}$ and we have $\partial B=$
$\partial S^{3}=\varnothing$. Now, as it is well known (see, e.g.,
\cite{nakahara}), the third de Rham cohomology group of $\mathbb{R\times}%
S^{3}$ is $H^{3}\left(  \mathbb{R\times}S^{3}\right)  =H^{3}\left(
S^{3}\right)  =\mathbb{R}$. Since this group is non trivial it follows that in
such manifold closed forms are not exact. Then from Eq.(\ref{7.10.22}) it did
not follow the validity of an equation analogous to Eq.(\ref{7.10.21}). So, in
that case an equation like Eq.(\ref{7.10.22.2}) cannot even be written.

Despite that commentary, keep in mind that in Einstein's theory the `energy'
of a closed universe\footnote{Note that if we suppose that the universe
contains spinor fields, as we indeed did, then it must be a spin manifold,
i.e., it is parallelizable according to Geroch's theorem \cite{geroch}.}
supposed to be given by Eq.(\ref{7.10.22.3'}) is indeed zero, since in that
theory the $3$-forms $(\star\mathcal{T}^{\mathbf{a}}+\star t^{\mathbf{a}})$
are indeed exact (see Eq.(\ref{7.10.21})). This means that accepting
$t^{\mathbf{a}}$ as the energy-momentum $1$-form fields of the gravitational
field, it follows that gravitational energy must be \textit{negative} in a
closed universe.

\textbf{(iii)} But, is the above formalism a consistent one? Given a
coordinate chart \ with "Cartesian" like coordinates $\{x^{\mu}\}$ of the
atlas of $M$, with some algebra (left as exercise to the reader) one can show
that for a gravitational model represented by a diagonal asymptotic flat
metric\footnote{A metric is said to be asymptotically flat in given
coordinates, if $g_{\mu\nu}=n_{\mu\nu}(1+\mathrm{O}\left(  r^{-k}\right)  )$,
with $k=2$ or $k=1$ depending on the author. See, e.g., \cite{schoenyau1,
schoenyau2,wald}.}, the inertial mass $E=m_{\mathbf{I}}$ is given by
\begin{equation}
m_{\mathbf{I}}=-\frac{1}{16\pi}\lim_{r\rightarrow\infty}%
{\displaystyle\int\limits_{\partial B}}
\frac{x_{i}}{r}\frac{\partial}{\partial x^{j}}(g_{11}g_{22}g_{33}g^{ij}%
)r^{2}d\Omega, \label{7.10.22.4}%
\end{equation}
where $\partial B=S^{2}(r)$ is a $2$-sphere of radius $r$, $g_{ij}x^{j}=x_{i}$
and $d\Omega$ is the element of solid angle. If we apply Eq.(\ref{7.10.22.4})
to calculate, e.g., the energy of the Schwarzschild space time\footnote{For a
Schwarzschild spacetime we have $g=\left(  1-\frac{2m}{r}\right)  dt\otimes
dt-\left(  1-\frac{2m}{r}\right)  ^{-1}dr\otimes dr-r^{2}(d\theta\otimes
d\theta+\sin^{2}\theta d\varphi\otimes d\varphi)$.} generate by a
gravitational mass $m$, we expect to have one unique and unambiguous result,
namely $m_{\mathbf{I}}=m$.

However, as shown in details, e.g., in \cite{boro} the calculation of $E$
depends on the spatial coordinate system naturally adapted to the reference
frame $Z=\frac{1}{\sqrt{\left(  1-\frac{2m}{r}\right)  }}\frac{\partial
}{\partial t}$ , even if these coordinates produce asymptotically flat
metrics. Then, even if in one given chart we may obtain $m_{\mathbf{I}}=m$
there are others where $m_{\mathbf{I}}\neq m$!\footnote{This observation is
true even if we use the so called ADM formalism \cite{adm}. To be more
precise,let us recall that we have a well defined ADM energy only ifthe fall
off rate of the metric is in the interval $1/2<k<1$. For details, see
\cite{murchada}.}

Moreover, note also that, as shown above, for a closed universe Einstein's
theory implies on general grounds (once we accept that the $t^{\mathbf{a}}$
describes the energy-momentum distribution of the gravitational field) that
$m_{\mathbf{I}}=0$. This result --- it is important to quote --- does not
contradict the so called ``positive mass theorems'' of, e.g., references
\cite{schoenyau1,schoenyau2,witten}, because those theorems refer to the total
energy of an isolated system. A system of that kind is supposed to be modelled
by a Lorentzian spacetime having a spacelike, asymptotically Euclidean
hypersurface.\footnote{The proof also uses as hypothesis the so called energy
dominance condition \cite{hawellis}.} However, we emphasize, although the
energy results positive, its value is not unique, since depends on the
asymptotically flat coordinates chosen to perform the calculations, as it is
clear from the elementary example of the Schwarzschild field commented above
and detailed in \cite{boro}.

In a book written in 1970, Davis \cite{davis} said:

\begin{quote}
{\footnotesize ``Today, some 50 years after the development of Einstein's
generally covariant field theory it appears that no general agreement
regarding the proper formulation of the conservation laws has been reached.''
}
\end{quote}

Well, we hope that the reader has been convinced that the fact is: there are
\textit{in general} no conservation laws of energy-momentum in General
Relativity. Moreover, all discourses (based on Einstein's equivalence
principle)\footnote{Like, e.g., in \cite{anderson,penrose05,mtw} and many
other textbooks. It is worth to quote here that, at least, Anderson
\cite{anderson} explicitly said: " In an interaction that involves the
gravitational field a system can lose energy without this energy being
transmitted to the gravitational field."} concerning the use of pseudo-energy
momentum tensors as \textit{reasonable} descriptions of energy and momentum of
gravitational fields in Einstein's theory are not convincing.

And, at this point it is better to quote page 98 of Sachs\&Wu \cite{sawu}:

\begin{quote}
{\footnotesize ``As mentioned in section 3.8, conservation laws have a great
predictive power. It is a shame to lose the special relativistic total energy
conservation law (Section 3.10.2) in general relativity. Many of the attempts
to resurrect it are quite interesting; many are simply garbage.'' }
\end{quote}

In GRT---we already said---every gravitational field is modelled (module
diffeomorphisms and according to present wisdom) by a Lorentzian spacetime. In
that particular case, when this spacetime structure admits a \textit{timelike}
Killing vector, we may formulate a law of energy conservation for the matter
fields. Also, if the Lorentzian spacetime admits three linearly independent
\textit{spacelike} Killing vectors, we have a law of conservation of momentum
for the matter fields.

This follows at once from the theory developed in the previous section.
Indeed, in the \textit{particular} case of General Relativity, the Lagrangian
density of the matter field is not supposed to be explicitly dependent on the
$\omega_{\mathbf{b}}^{\mathbf{a}}$. Then, $\frac{\partial\mathcal{L}_{m}%
}{\partial\omega_{\mathbf{b}}^{\mathbf{a}}}=0$ in Eq.(\ref{7.40}) and writing
$\mathcal{T}(\xi)=\xi^{\mu}\mathcal{T}_{\mu}$, it becomes $d\star
\mathcal{T}(\xi)=0$, or
\begin{equation}
\delta\mathcal{T}(\xi)=0. \label{7.61}%
\end{equation}
The crucial fact to have in mind here is that a general Lorentzian spacetime,
does \emph{not} admit such Killing vectors in general, as it is the case,
e.g., of the popular Friedmann-Robertson-Walker expanding universes models.

At present, the authors know only one possibility of resurrecting a
\textit{trustworthy} conservation law of energy-momentum valid in all
circumstances in a theory of the gravitational field that \textit{resembles}
General Relativity (in the sense of keeping Einstein's equation). It consists
in reinterpreting that theory as a field theory in flat Minkowski spacetime.
Theories of this kind have been proposed in the past by, e.g.,
Feynman~\cite{Feynman}, Schwinger~\cite{schwinger}, Thirring~\cite{thirring0}
and Weinberg~\cite{weinberg1} among others and have been extensively studied
by Logunov and collaborators in a series of papers summarized in the
monographs \cite{logunov1,logunov2} and also in \cite{roqui,rodoliv2006}.

\section{Is there any Angular Momentum Conservation law in the GRT}

If the $\{\theta^{\mathbf{a}}\}$ and the $\{\omega_{\mathbf{b}}^{\mathbf{a}%
}\}$ are varied independently in the Einstein-Hilbert Lagrangian then, as it
is easy to verify we get the additional field equation\footnote{$\theta
^{\mathbf{ab}}=\theta^{\mathbf{a}}\wedge\theta^{\mathbf{b}}$ is known
\cite{szabados} as Bramson \cite{bramson} superpotential}%
\begin{equation}
\mathbf{D\star}\theta^{\mathbf{ab}}=\mathbf{J}^{\mathbf{ab}} \label{7.10.21a}%
\end{equation}

From this equation we get immediately
\begin{equation}
d\mathbf{\star}\theta_{\mathbf{b}}^{\mathbf{a}}=\mathbf{J}_{\mathbf{b}%
}^{\mathbf{a}}-\omega_{\mathbf{b}}^{\mathbf{c}}\wedge\mathbf{\star}%
\theta_{\mathbf{c}}^{\mathbf{a}}+\mathbf{\star}\theta_{\mathbf{b}}%
^{\mathbf{c}}\wedge\omega_{\mathbf{c}}^{\mathbf{a}}%
\end{equation}
and one is tempted to define $\mathbf{S}_{\mathbf{b}}^{\mathbf{a}}%
=(\omega_{\mathbf{b}}^{\mathbf{c}}\wedge\mathbf{\star\theta}_{\mathbf{c}%
}^{\mathbf{a}}+\mathbf{\star\theta}_{\mathbf{b}}^{\mathbf{c}}\wedge
\omega_{\mathbf{c}}^{\mathbf{a}})$ as the density of \textit{spin} angular
momentum of the gravitational field and the (total) angular momentum of the
system as%
\begin{equation}
\mathbf{L}_{\mathbf{b}}^{\mathbf{a}}:=%
{\displaystyle\int\nolimits_{S^{2}}}
\mathbf{\star\theta}_{\mathbf{b}}^{\mathbf{a}}. \label{angular}%
\end{equation}

This definition, of course, has the same problems as the definition of energy
in the GRT because $\mathbf{S}_{\mathbf{b}}^{\mathbf{a}}$ \ is gauge dependent.

\section{Conservation Laws in the Teleparallel Equivalent of General
Relativity}

We observe that recently it was claimed \cite{deandrade} a valid way of
formulating a genuine energy-momentum conservation law in a theory equivalent
to General Relativity. In that theory, the so-called \textit{teleparallel}
equivalent of General Relativity theory \cite{maluf}, spacetime is
teleparallel (or Weitzenb\"{o}ck), i.e., has a metric compatible connection
with non zero torsion and with null curvature\footnote{In fact, formulation of
teleparallel equivalence of General Relativity is a subject with an old
history. See, e.g., \cite{haya,kop,mol,miel,tung}.}. However, the claim of
\cite{deandrade} is investigated in more detail below. Indeed, we have two
important comments (\textbf{a}) and (\textbf{b}) concerning this
issue.\medskip

(\textbf{a) }First, it must be clear that the mathematical structure of the
teleparallel equivalent of General Relativity consists in the introduction of:
(i) a bilinear form (a deformed metric tensor) \texttt{$\mathbf{g}$ }%
$=\eta_{\mathbf{ab}}\theta^{\mathbf{a}}\otimes\theta^{\mathbf{b}}$ and (ii) a
teleparallel connection in a manifold $M\simeq\mathbb{R}^{4}$ (the same which
appears in the Minkowski spacetime structure). Indeed, taking advantage of the
the discussion of the previous sections, we can present that theory with a
cosmological constant term as follows. Start with the Lagrangian density
$\mathcal{L}^{\prime}=\mathcal{L}_{g}+\mathcal{L}_{m}$, where\footnote{Field
equations in Maxwell like form for $F^{\mathbf{a}}=d\theta^{\mathbf{a}}$ are
presented in \cite{rodrigues2006}.}%
\[
\mathcal{L}_{g}^{\prime}=-\frac{1}{2}d\theta^{\mathbf{a}}\wedge\star
d\theta_{\mathbf{a}}+\frac{1}{2}\delta\theta^{\mathbf{a}}\wedge\star
\delta\theta_{\mathbf{a}}+\frac{1}{4}\left(  d\theta^{\mathbf{a}}\wedge
\theta_{\mathbf{a}}\right)  \wedge\star\left(  d\theta^{\mathbf{b}}%
\wedge\theta_{\mathbf{b}}\right)  +\frac{1}{2}m^{2}\theta_{\mathbf{a}}%
\wedge\star\theta^{\mathbf{a}}%
\]
and write it (after some algebraic manipulations) as%
\begin{align}
\mathcal{L}_{g}^{\prime}  &  =-\frac{1}{2}d\theta^{\mathbf{a}}\wedge
\star\left[  d\theta_{\mathbf{a}}-\theta_{\mathbf{a}}\wedge(\theta
_{\mathbf{b}}\lrcorner d\theta_{\mathbf{b}})+\frac{1}{2}\star\theta
_{\mathbf{a}}\wedge\star(d\theta^{\mathbf{b}}\wedge\theta_{\mathbf{b}%
})\right]  +\frac{1}{2}m^{2}\theta_{\mathbf{a}}\wedge\star\theta^{\mathbf{a}%
}\nonumber\\
&  =-\frac{1}{2}d\theta^{\mathbf{a}}\wedge\star(^{(1)}d\theta_{\mathbf{a}%
}-2^{(2)}d\theta_{\mathbf{a}}-\frac{1}{2}\text{ }^{(3)}d\theta_{\mathbf{a}%
})+\frac{1}{2}m^{2}\theta_{\mathbf{a}}\wedge\star\theta^{\mathbf{a}},
\label{tele1}%
\end{align}
where
\begin{align}
d\theta^{\mathbf{a}}  &  =^{{{}(1)}}d\theta^{\mathbf{a}}+^{{{}(2)}}%
d\theta^{\mathbf{a}}+^{{{}(3)}}d\theta^{\mathbf{a}},\nonumber\\
^{{{}(1)}}d\theta^{\mathbf{a}}  &  =d\theta^{\mathbf{a}}-^{{{}(2)}}%
d\theta^{\mathbf{a}}-^{{{}(3)}}d\theta^{\mathbf{a}},\nonumber\\
^{{{}(2)}}d\theta^{\mathbf{a}}  &  =\frac{1}{3}\theta^{\mathbf{a}}%
\wedge(\theta^{\mathbf{b}}\lrcorner d\theta_{\mathbf{b}}),\nonumber\\
^{{{}(3)}}d\theta^{\mathbf{a}}  &  =-\frac{1}{3}\star(\theta^{\mathbf{a}%
}\wedge\star(d\theta^{\mathbf{b}}\wedge\theta_{\mathbf{b}})). \label{tele2}%
\end{align}
Next introduce a teleparallel connection by declaring that the cobasis
$\{\theta^{\mathbf{a}}\}$ fixes the parallelism, i.e., we define the torsion
$2$-forms by
\begin{equation}
\Theta^{\mathbf{a}}:=d\theta^{\mathbf{a}}, \label{tele3}%
\end{equation}
and $\mathcal{L}_{g}^{\prime}$ becomes%
\begin{equation}
\mathcal{L}_{g}^{\prime}=-\frac{1}{2}\Theta^{\mathbf{a}}\wedge\star\left(
^{{{}(1)}}\Theta^{\mathbf{a}}-2^{({{}2)}}\Theta^{\mathbf{a}}-\frac{1}{2}\text{
}^{{{}(3)}}\Theta^{\mathbf{a}}\text{ }\right)  +\frac{1}{2}m^{2}%
\theta_{\mathbf{a}}\wedge\star\theta^{\mathbf{a}}, \label{tele4}%
\end{equation}
where $^{{{}(1)}}\Theta^{\mathbf{a}}=^{{{}(1)}}d\theta^{\mathbf{a}}$,
$^{({{}2)}}\Theta^{\mathbf{a}}=^{{{}(2)}}d\theta^{\mathbf{a}}$ and $^{{{}(3)}%
}\Theta^{\mathbf{a}}=^{{{}(3)}}d\theta^{\mathbf{a}}$, called
\ \textit{tractor} (four components), \textit{axitor} (four components) and
\textit{tentor} (sixteen components) are the irreducible components of the
tensor torsion under the action of \textrm{SO}$_{1,3}^{e}$.\medskip

(\textbf{b}) Recalling the results of the previous sections we now show that
even if the metric of a given teleparallel spacetime has some Killing vector
fields there are genuine conservation laws involving only the energy-momentum
and angular momentum tensors of \textit{matter }only if some additional
condition is satisfied\textit{.} Indeed, in the teleparallel basis where
$\nabla_{\mathbf{e}_{\mathbf{a}}}\mathbf{e}_{\mathbf{b}}=0$ and $[\mathbf{e}%
_{\mathbf{m}},\mathbf{e}_{\mathbf{n}}]=c_{\mathbf{mn}}^{\mathbf{a}}%
\mathbf{e}_{\mathbf{a}}$ we have that the torsion $2$-forms satisfies
\begin{equation}
\Theta^{\mathbf{a}}=d\theta^{\mathbf{a}}=-\frac{1}{2}c_{\mathbf{mn}%
}^{\mathbf{a}}\theta^{\mathbf{m}}\wedge\theta^{\mathbf{n}}=\frac{1}%
{2}T_{\mathbf{mn}}^{\mathbf{a}}\theta^{\mathbf{m}}\wedge\theta^{\mathbf{n}}.
\label{7.61a}%
\end{equation}
Then, recalling once again that $\pounds _{\xi}(d\theta^{\mathbf{a}})=$
$d(\pounds _{\xi}\theta^{\mathbf{a}})=d(\varkappa_{\mathbf{b}}^{\mathbf{a}%
}\theta^{\mathbf{b}})$ and Eq.(\ref{7.42a}) we can use Eq.(\ref{7.45}) (which
express the condition $\pounds _{\xi}\Theta=\mathbf{0}$) to write%
\begin{equation}
d(\varkappa_{\mathbf{b}}^{\mathbf{a}}\theta^{\mathbf{b}})=\varkappa
_{\mathbf{b}}^{\mathbf{a}}d\theta^{\mathbf{b}}, \label{7.61'}%
\end{equation}
which implies
\begin{equation}
d\varkappa_{\mathbf{b}}^{\mathbf{a}}\wedge\theta^{\mathbf{b}}=0. \label{7.61b}%
\end{equation}
Then, Eq.(\ref{7.61b}) is satisfied only if the torsion tensor of the
teleparallel spacetime satisfy the following differential equation:
\begin{equation}
T_{\mathbf{bd}}^{\mathbf{m}}\mathbf{e}_{\mathbf{m}}(\xi^{\mathbf{a}%
})+\mathbf{e}_{\mathbf{d}}(\xi^{\mathbf{m}}T_{\mathbf{bm}}^{\mathbf{a}%
})-\mathbf{e}_{\mathbf{b}}(\xi^{\mathbf{m}}T_{\mathbf{dm}}^{\mathbf{a}})=0.
\label{7.62}%
\end{equation}

Of course, Eq.(\ref{7.62}) is in general \textit{not} satisfied for a vector
field $\xi$ that is simply a Killing vector of \texttt{$\mathbf{g}$%
}$\mathtt{\mathbf{.}}$ This means that in the teleparallel equivalent of
General Relativity even if there are Killing vector fields, this in general do
not warrant that there are conservation laws as in Eq.(\ref{7.40}) involving
\textit{only} the energy and angular momentum tensors of \textit{matter}.

Next, we remark that from $\mathcal{L}_{g}^{\prime}$ we get as field equations
(in an arbitrary basis, not necessarily the teleparallel one) satisfied by the
gravitational field the Eq.(\ref{7.10.21}), i.e.,
\begin{equation}
-d\star\mathcal{S}^{\mathbf{a}}=\star\mathcal{T}^{\mathbf{a}}+\text{ }%
\star\mathfrak{t}^{\mathbf{a}}, \label{war33}%
\end{equation}
with%
\[
\star\mathfrak{t}^{\mathbf{a}}=\star t^{\mathbf{a}}+m^{2}\star\theta
^{\mathbf{a}}%
\]
and $\mathcal{S}^{\mathbf{a}}$ and $t^{\mathbf{a}}$ given in Eq.(\ref{7.10.17}%
) where it must also be taken into account that in the teleparallel equivalent
of General Relativity and using the teleparallel basis the Levi-Civita
connection $1$-forms $\omega_{\mathbf{b}}^{\mathbf{a}}$ there must be
substituted by $-\kappa_{\mathbf{b}}^{\mathbf{a}}$, with
\begin{align}
\kappa^{\mathbf{cd}}  &  =-\frac{1}{2}\left[  \theta^{\mathbf{d}}\lrcorner
d\theta^{\mathbf{c}}-\theta^{\mathbf{c}}\lrcorner d\theta^{\mathbf{d}}+\left(
\theta^{\mathbf{c}}\lrcorner(\theta^{\mathbf{d}}\lrcorner d\theta_{\mathbf{a}%
})\right)  \theta^{\mathbf{a}}\right] \nonumber\\
&  =-\frac{1}{2}\left[  \theta^{\mathbf{d}}\lrcorner\Theta^{\mathbf{c}}%
-\theta^{\mathbf{c}}\lrcorner\Theta^{\mathbf{d}}+\left(  \theta^{\mathbf{c}%
}\lrcorner(\theta^{\mathbf{d}}\lrcorner\Theta_{\mathbf{a}})\right)
\theta^{\mathbf{a}}\right]  ,
\end{align}
where $\kappa_{\mathbf{b}}^{\mathbf{a}}=K_{\mathbf{bc}}^{\mathbf{a}}%
\theta^{\mathbf{c}}$, with $K_{\mathbf{bc}}^{\mathbf{a}}$ the components of
the so called contorsion tensor. We have,%
\begin{equation}
\star t^{\mathbf{c}}=\frac{1}{2}\kappa_{\mathbf{ab}}\wedge\lbrack
\kappa_{\mathbf{d}}^{\mathbf{c}}\wedge\star(\theta^{\mathbf{a}}\wedge
\theta^{\mathbf{b}}\wedge\theta^{\mathbf{c}})+\kappa_{\mathbf{d}}^{\mathbf{b}%
}\wedge\star(\theta^{\mathbf{a}}\wedge\theta^{\mathbf{b}}\wedge\theta
^{\mathbf{c}})].
\end{equation}
Under a change of gauge, $\theta^{\mathbf{a}}\mapsto\theta^{\prime\mathbf{a}%
}=U\theta^{\mathbf{a}}U=\Lambda_{\mathbf{b}}^{\mathbf{a}}\theta^{\mathbf{b}}$
$(U\in\sec\mathrm{Spin}_{1,3}^{e}(M)\hookrightarrow\mathcal{C}\ell
(M,\mathtt{g})$, $\Lambda_{\mathbf{b}}^{\mathbf{a}}(x)\in\mathrm{SO}_{1,3}%
^{e}$, $\forall$ $x\in M$), we have that $\Theta^{\mathbf{a}}\mapsto
\Theta^{\prime\mathbf{a}}=\Lambda_{\mathbf{b}}^{\mathbf{a}}\Theta^{\mathbf{b}%
}$.\ It follows that the $\mathfrak{t}_{\mathbf{b}}^{\mathbf{a}}$, which are
the components of the energy-momentum 1-forms $\mathfrak{t}^{\mathbf{a}%
}=\mathfrak{t}_{\mathbf{b}}^{\mathbf{a}}\theta^{\mathbf{b}}$ defines a tensor field.

We then conclude that for each gravitational field modelled by a particular
teleparallel spacetime, if the cosmological term is null or not there is a
conservation law of energy-momentum for the coupled system of the matter field
and the gravitational field which is represented by that \textit{particular}
teleparallel spacetime. Although the existence of such a conservation law in
the teleparallel spacetime is a satisfactory fact with respect of the usual
formulation of the gravitational theory where gravitational fields are
modelled by Lorentzian spacetimes and where genuine conservation laws (in
general) does not exist because in that theory the components of
$t^{\mathbf{a}}$ defines only a pseudo-tensor, we cannot forget observation
(\textbf{a}): the teleparallel equivalent of General Relativity consists in
the introduction of: (i) a bilinear form (a deformed metric tensor)
\texttt{$\mathbf{g}$ }$=\eta_{\mathbf{ab}}\theta^{\mathbf{a}}\otimes
\theta^{\mathbf{b}}$ and (ii) a teleparallel connection in the manifold
$M\simeq\mathbb{R}^{4}$ of Minkowski spacetime structure. The crucial
ingredient is still the Einstein-Hilbert Lagrangian density.

Finally we must remark that if we insist in working with a teleparallel
spacetime we lose in general the other six genuine angular momentum
conservation laws which always hold in Minkowski spacetime. Indeed, we do not
obtain in general even the chart dependent angular momentum `conservation' law
of GRT. The reason is that if we write the equivalent of Eq.(\ref{7.10.21}) in
a chart $(U,\varphi)$ with coordinates $\{x^{\mu}\}$ for $U\subset M$ we did
not get in general that $dx^{\mu}\wedge\star t^{\nu}=dx^{\nu}\wedge\star
t^{\mu}$, which as well known is necessary in order to have a chart dependent
angular momentum conservation law \cite{thirring}.

\section{Conclusions}

We recall that the problem of the conservation laws of energy-momentum and
angular momentum in GRT occupied the mind of many people since Einstein
\cite{einstein} introduced the so called energy-momentum pseudo-tensor in
1916. Besides those papers that already have been quoted above it is worth to
cite also
\cite{adm,brown,east,east2,franferr,freud,komar1,komar2,komar3,komar4,komar5,mingu,moller,moller2,nester,trautman0}%
, which---summed with the quote of \cite{sawu} presented in Section~5---have
been the inspiration for the present work, where we recalled (a) under which
conditions there exists genuine conservation laws of energy-momentum and
angular momentum involving only the matter fields on a general RCST and (b)
under which conditions there exists genuine conservation laws involving both
the energy-momentum and angular momentum tensors of the matter and the
gravitational field, when this latter concept can be rigorously defined.

Using a Clifford bundle formalism it was shown that in case (a) contrary to
the case of GRT the simply existence of Killing vector fields is not enough,
since a new additional condition must hold. Some examples are presented in Appendix~B.

Concerning case (b) our conclusion is that genuine laws involving both the
energy-momentum and angular momentum tensors of the matter and the
gravitational field exist only in a field theory of the gravitational field
formulated in Minkowski spacetime. We analyzed also a particular case of a
RCST theory, namely the so called teleparallel equivalent of GRT
\cite{maluf,maluf2,deandrade}. In that theory a genuine conservation law of
energy-momentum is obtained through the introduction of a teleparallel
connection, needed to restore active Local Lorentz invariance\footnote{We
recall that recently it has been shown that imposition of active local Lorentz
invariance in theories containing, e.g., spinor fields implies in an
equivalence of spacetimes with different curvatures and/or different torsion
tensors \cite{roldaowa,rodroldao}.}. However, in the teleparallel equivalent
of GRT, it is not possible (in general) to formulate even a chart dependent
conservation law for the angular momentum of matter or for both the matter and
gravitational fields. Due to this fact, in our opinion it cannot be considered
more general than a formulation of a theory of the gravitational field which
uses a deformation tensor in Minkowski spacetime structure
\cite{roqui,rodrigues2006}, where the introduction of general connections are
not needed.

\appendix

\section{Clifford and Spin-Clifford Bundles}

Let $\mathcal{M}=(M,\mathbf{g},\mathbf{\nabla},\tau_{\mathbf{g}},\uparrow)$ be
an arbitrary Riemann-Cartan spacetime.\ The quadruple $(M,\mathbf{g}%
,\tau_{\mathbf{g}},\uparrow)$ denotes a four-dimensional time-oriented and
space-oriented Lorentzian manifold. This means that $\mathbf{g}\in\sec
T_{2}^{0}M$ is a Lorentzian metric of signature (1,3), $\tau_{\mathbf{g}}%
\in\sec\bigwedge{}^{4}(T^{\ast}M)$ and $\uparrow$ is a time-orientation (see
details, e.g., in \cite{sawu}). Here, $T^{\ast}M$ [$TM$] is the cotangent
[tangent] bundle. $T^{\ast}M=\cup_{x\in M}T_{x}^{\ast}M$, $TM=\cup_{x\in
M}T_{x}M$, and $T_{x}M\simeq T_{x}^{\ast}M\simeq\mathbb{R}^{1,3}$, where
$\mathbb{R}^{1,3}$ is the Minkowski vector space\footnote{Not to be confused
with Minkowski spacetime \cite{sawu}.}. $\mathbf{\nabla}$ is an arbitrary
\ metric compatible connection, i.e.\/, $\mathbf{\nabla g}=0$, but in general,
$\mathbf{R}^{\mathbf{\nabla}}\neq0$, $\Theta^{\mathbf{\nabla}}\neq0$,
$\mathbf{R}^{\mathbf{\nabla}}$ and $\Theta^{\mathbf{\nabla}}$ being
respectively the curvature and torsion tensors of the connection
$\mathbf{\nabla}$. \ When $\mathbf{R}^{\mathbf{\nabla}}\neq0$, $\mathbf{T}%
(\mathbf{\nabla})\neq0$, $\mathcal{M}$ is called a \textit{Riemann-Cartan
spacetime}. When $\mathbf{R}^{\mathbf{\nabla}}=0$, $\Theta^{\mathbf{\nabla}%
}\neq0$, $\mathcal{M}$ is called a \textit{teleparallel }(or\textit{
Weintzb\"{o}ck})\textit{ spacetime. }For a Lorentzian spacetime the connection
is the Levi-Civita connection $D$ of $\mathbf{g}$ for which $\mathbf{R}%
^{D}\neq0$, $\Theta^{D}=0$. \ Minkowski spacetime is the particular case of a
Lorentzian spacetime for which $\mathbf{R}^{D}=0$, $\Theta^{D}=0$, and
$M\simeq\mathbb{R}^{4}$. Let $\mathtt{g}\in\sec T_{0}^{2}M$ be the metric of
the \textit{cotangent bundle}. The Clifford bundle of differential forms
$\mathcal{C}\!\ell(M,\mathtt{g})$ is the bundle of algebras, i.e.,
$\mathcal{C}\!\ell(M,\mathtt{g})=\cup_{x\in M}\mathcal{C}\!\ell(T_{x}^{\ast
}M,\mathtt{g})$, where $\forall x\in M$, $\mathcal{C}\!\ell(T_{x}^{\ast
}M,\mathtt{g})=\mathbb{R}_{1,3}$, the so called \emph{spacetime} \emph{algebra
}\cite{rod04}. Recall also that $\mathcal{C}\!\ell(M,\mathtt{g})$ is a vector
bundle associated to the \emph{orthonormal frame bundle}, i.e., $\mathcal{C}%
\!\ell(M,\mathtt{g})$ $=P_{\mathrm{SO}_{(1,3)}^{e}}(M)\times_{\mathrm{Ad}%
}\mathcal{C}l_{1,3}$ \cite{lawmi,moro}. For any $x\in M$, $\mathcal{C}%
\!\ell(T_{x}^{\ast}M,\left.  \mathtt{g}\right\vert _{x})$ as a linear space
over the real field $\mathbb{R}$ is isomorphic to the Cartan algebra
$\bigwedge T_{x}^{\ast}M$ of the cotangent space. $\bigwedge T_{x}^{\ast
}M=\oplus_{k=0}^{4}\bigwedge^{k}T_{x}^{\ast}M$, where $\bigwedge^{k}%
T_{x}^{\ast}M$ is the $\binom{4}{k}$-dimensional space of $k$-forms. Then,
sections of $\mathcal{C}\!\ell(M,\mathtt{g})$ can be represented as a sum of
non homogeneous differential forms, that will be called Clifford (multiform)
fields. Let $\{\mathbf{e}_{\mathbf{a}}\}\in\sec P_{\mathrm{SO}_{(1,3)}^{e}%
}(M)$ (the frame bundle) be an orthonormal basis for $TU\subset TM$, i.e.,
$\mathtt{g}(\mathbf{e}_{\mathbf{a}},\mathbf{e}_{\mathbf{a}})=\eta
_{\mathbf{ab}}=\mathrm{diag}(1,-1,-1,-1)$. Let $\theta^{\mathbf{a}}\in
\sec\bigwedge^{1}T^{\ast}M\hookrightarrow\sec\mathcal{C}\!\ell(M,\mathtt{g})$
($\mathbf{a}=0,1,2,3$) be such that the set $\{\theta^{\mathbf{a}}\}$ is the
dual basis of $\{\mathbf{e}_{\mathbf{a}}\}$.

\subsection{Clifford Product}

The fundamental \emph{Clifford product} (in what follows to be denoted by
juxtaposition of symbols) is generated by $\theta^{\mathbf{a}}\theta
^{\mathbf{b}}+\theta^{\mathbf{b}}\theta^{\mathbf{a}}=2\eta^{\mathbf{ab}}$ and
if $\mathcal{C}\in\sec\mathcal{C}\!\ell(M,\mathtt{g})$ we have%

\begin{equation}
\mathcal{C}=s+v_{\mathbf{a}}\theta^{\mathbf{a}}+\frac{1}{2!}f_{\mathbf{ab}%
}\theta^{\mathbf{a}}\theta^{\mathbf{b}}+\frac{1}{3!}t_{\mathbf{abc}}%
\theta^{\mathbf{a}}\theta^{\mathbf{b}}\theta^{\mathbf{c}}+p\theta^{5}\;,
\label{3}%
\end{equation}
where $\tau_{\mathtt{g}}=\theta^{5}=\theta^{0}\theta^{1}\theta^{2}\theta^{3}$
is the volume element and $s$, $v_{\mathbf{a}}$, $f_{\mathbf{ab}}$,
$t_{\mathbf{abc}}$, $p\in\sec\bigwedge^{0}T^{\ast}M\hookrightarrow
\sec\mathcal{C}\!\ell(M,\mathtt{g})$.

For $A_{r}\in\sec\bigwedge^{r}T^{\ast}M\hookrightarrow\sec\mathcal{C}%
\!\ell(M,\mathtt{g}),B_{s}\in\sec\bigwedge^{s}T^{\ast}M\hookrightarrow
\sec\mathcal{C}\!\ell(M,\mathtt{g})$ we define the \emph{exterior product} in
$\mathcal{C}\!\ell(M,\mathtt{g})$ \ ($\forall r,s=0,1,2,3)$ by
\begin{equation}
A_{r}\wedge B_{s}=\langle A_{r}B_{s}\rangle_{r+s}, \label{5}%
\end{equation}
where $\langle\;\;\rangle_{k}$ is the component in $\bigwedge^{k}T^{\ast}M$ of
the Clifford field. Of course, $A_{r}\wedge B_{s}=(-1)^{rs}B_{s}\wedge A_{r}$,
and the exterior product is extended by linearity to all sections of
$\mathcal{C}\!\ell(M,\mathtt{g})$.

Let $A_{r}\in\sec\bigwedge^{r}T^{\ast}M\hookrightarrow\sec\mathcal{C}%
\!\ell(M,\mathtt{g}),B_{s}\in\sec\bigwedge^{s}T^{\ast}M\hookrightarrow
\sec\mathcal{C}\!\ell(M,\mathtt{g})$. We define a \emph{scalar product
}in\emph{\ }$\mathcal{C}\!\ell(M,\mathtt{g})$ (denoted by $\cdot$) as follows:

(i) For $a,b\in\sec\bigwedge^{1}(T^{\ast}M)\hookrightarrow\sec\mathcal{C}%
\!\ell(M,\mathtt{g}),$%
\begin{equation}
a\cdot b=\frac{1}{2}(ab+ba)=\mathtt{g}(a,b). \label{4}%
\end{equation}

(ii) For $A_{r}=a_{1}\wedge...\wedge a_{r},B_{r}=b_{1}\wedge...\wedge b_{r}$,
$a_{i},b_{j}\in\sec\bigwedge^{1}T^{\ast}M\hookrightarrow\sec\mathcal{C}%
\!\ell(M,\mathtt{g})$, $i,j=1,...,r,$
\begin{align}
A_{r}\cdot B_{r}  &  =(a_{1}\wedge...\wedge a_{r})\cdot(b_{1}\wedge...\wedge
b_{r})\nonumber\\
&  =\left\vert
\begin{array}
[c]{lll}%
a_{1}\cdot b_{1} & .... & a_{1}\cdot b_{r}\\
.......... & .... & ..........\\
a_{r}\cdot b_{1} & .... & a_{r}\cdot b_{r}%
\end{array}
\right\vert . \label{6}%
\end{align}

We agree that if $r=s=0$, the scalar product is simply the ordinary product in
the real field.

Also, if $r\neq s$, then $A_{r}\cdot B_{s}=0$. Finally, the scalar product is
extended by linearity for all sections of $\mathcal{C}\!\ell(M,\mathtt{g})$.

For $r\leq s$, $A_{r}=a_{1}\wedge...\wedge a_{r}$, $B_{s}=b_{1}\wedge...\wedge
b_{s\text{ }}$, we define the \textit{left contraction} $\lrcorner
:(A_{r},B_{s})\mapsto A_{r}\mathbin\lrcorner B_{s}$ by
\begin{equation}
A_{r}\mathbin\lrcorner B_{s}=%
{\displaystyle\sum\limits_{i_{1}\,<...\,<i_{r}}}
\epsilon^{i_{1}...i_{s}}(a_{1}\wedge...\wedge a_{r})\cdot(b_{_{i_{1}}}%
\wedge...\wedge b_{i_{r}})^{\sim}b_{i_{r}+1}\wedge...\wedge b_{i_{s}}\label{7}%
\end{equation}
where $\sim$ is the reverse mapping (\emph{reversion}) defined by
\begin{align}
\symbol{126} &  :\sec\mathcal{C}\!\ell(M,\mathtt{g})\rightarrow\sec
\mathcal{C}\!\ell(M,\mathtt{g}),\nonumber\\
\tilde{A} &  =%
{\displaystyle\sum\limits_{p=0}^{4}}
\text{ }\tilde{A}_{p}=%
{\displaystyle\sum\limits_{p=0}^{4}}
(-1)^{\frac{1}{2}k(k-1)}A_{p},\nonumber\\
A_{p} &  \in\sec%
{\displaystyle\bigwedge\nolimits^{p}}
T^{\ast}M\hookrightarrow\sec\mathcal{C}\!\ell(M,\mathtt{g}).
\end{align}
We agree that for $\alpha,\beta\in\sec\bigwedge^{0}T^{\ast}M$ the contraction
is the ordinary (pointwise) product in the real field and that if $\alpha
\in\sec\bigwedge^{0}T^{\ast}M$, $A_{r}\in\sec\bigwedge^{r}T^{\ast}M,B_{s}%
\in\sec\bigwedge^{s}T^{\ast}M\hookrightarrow$ then $(\alpha A_{r}%
)\mathbin\lrcorner B_{s}=A_{r}\mathbin\lrcorner(\alpha B_{s})$. Left
contraction is extended by linearity to all pairs of sections of
$\mathcal{C}\!\ell(M,\mathtt{g})$, i.e., for $A,B\in\sec\mathcal{C}%
\!\ell(M,\mathtt{g})$%
\begin{equation}
A\mathbin\lrcorner B=\sum_{r,s}\langle A\rangle_{r}\mathbin\lrcorner\langle
B\rangle_{s},\quad r\leq s\label{9}%
\end{equation}

It is also necessary to introduce the operator of \emph{right contraction}
denoted by $\llcorner$. The definition is obtained from the one presenting the
left contraction with the imposition that $r\geq s$ and taking into account
that now if $A_{r}\in\sec\bigwedge^{r}T^{\ast}M,$ $B_{s}\in\sec\bigwedge
^{s}T^{\ast}M$ then $A_{r}\mathbin\llcorner(\alpha B_{s})=(\alpha
A_{r})\mathbin\llcorner B_{s}$. See also the third formula in Eq.(\ref{10}).

The main formulas used in this paper can be obtained from the following ones
\begin{align}
aB_{s}  &  =a\mathbin\lrcorner B_{s}+a\wedge B_{s},\text{ }B_{s}%
a=B_{s}\mathbin\llcorner a+B_{s}\wedge a,\nonumber\\
a\mathbin\lrcorner B_{s}  &  =\frac{1}{2}(aB_{s}-(-)^{s}B_{s}a),\nonumber\\
A_{r}\mathbin\lrcorner B_{s}  &  =(-)^{r(s-1)}B_{s}\mathbin\llcorner
A_{r},\nonumber\\
a\wedge B_{s}  &  =\frac{1}{2}(aB_{s}+(-)^{s}B_{s}a),\nonumber\\
A_{r}B_{s}  &  =\langle A_{r}B_{s}\rangle_{|r-s|}+\langle A_{r}%
\mathbin\lrcorner B_{s}\rangle_{|r-s-2|}+...+\langle A_{r}B_{s}\rangle
_{|r+s|}\nonumber\\
&  =\sum\limits_{k=0}^{m}\langle A_{r}B_{s}\rangle_{|r-s|+2k},\label{10}\\
A_{r}\cdot B_{r}  &  =B_{r}\cdot A_{r}=\tilde{A}_{r}\mathbin\lrcorner
B_{r}=A_{r}\mathbin\llcorner\tilde{B}_{r}=\langle\tilde{A}_{r}B_{r}\rangle
_{0}=\langle A_{r}\tilde{B}_{r}\rangle_{0},\nonumber
\end{align}
where $a\in\sec\bigwedge^{1}T^{\ast}M\hookrightarrow\sec\mathcal{C}%
\!\ell(M,\mathtt{g})$.

\subsubsection{Hodge Star Operator}

Let $\star$ be the Hodge star operator, i.e., the mapping
\[
\star:%
{\displaystyle\bigwedge\nolimits^{k}}
T^{\ast}M\rightarrow%
{\displaystyle\bigwedge\nolimits^{4-k}}
T^{\ast}M,\text{ }A_{k}\mapsto\mathop\star A_{k}%
\]
where for $A_{k}\in\sec\bigwedge^{k}T^{\ast}M\hookrightarrow\sec
\mathcal{C}\!\ell(M,\mathtt{g})$%
\begin{equation}
\lbrack B_{k}\cdot A_{k}]\tau_{g}=B_{k}\wedge\mathop\star A_{k},\forall
B_{k}\in\sec\bigwedge\nolimits^{k}T^{\ast}M\hookrightarrow\sec\mathcal{C}%
\!\ell(M,\mathtt{g}). \label{11a}%
\end{equation}
$\tau_{\mathtt{g}}=\theta^{5}\in\sec\bigwedge^{4}T^{\ast}M\hookrightarrow
\sec\mathcal{C}\!\ell(M,\mathtt{g})$ is a \emph{standard} volume element. Then
we can verify that
\begin{equation}
\mathop\star A_{k}=\widetilde{A}_{k}\theta^{5}. \label{11b}%
\end{equation}

\subsubsection{Dirac Operator}

Let $d$ and $\delta$ be respectively the differential and Hodge codifferential
operators acting on sections of $\mathcal{C}\!\ell(M,\mathtt{g})$. If
$A_{p}\in\sec\bigwedge^{p}T^{\ast}M\hookrightarrow\sec\mathcal{C}%
\!\ell(M,\mathtt{g})$, then $\delta A_{p}=(-1)^{p}\mathop\star^{-1}%
d\mathop\star A_{p}$, with $\star^{-1}\star=\mathrm{identity}$.

\begin{remark}
When there is necessity of specifying the metric field $\mathtt{\mathbf{g}}$
used in the definition of the Hodge star operator and the Hodge codifferential
operator we use the notations $\underset{\mathtt{\mathbf{g}}}{\star}$ and
$\underset{\mathtt{g}}{\delta}$.
\end{remark}

The Dirac operator acting on sections of $\mathcal{C}\!\ell(M,\mathtt{g})$
associated to a general metric compatible connection $\mathbf{\nabla}$ is the
invariant first order differential operator
\begin{equation}
{\mbox{\boldmath$\partial$}}^{rc}=\theta^{\mathbf{a}}\mathbf{\nabla
}_{\mathbf{e}_{\mathbf{a}}}, \label{12}%
\end{equation}
where $\{\mathbf{e}_{\mathbf{a}}\}$ is an arbitrary \emph{orthonormal basis}
for $TU\subset TM$ and $\{\theta^{\mathbf{b}}\}$ is a basis for $T^{\ast
}U\subset T^{\ast}M$ dual to the basis $\{\mathbf{e}_{\mathbf{a}}\}$, i.e.,
$\theta^{\mathbf{b}}(\mathbf{e}_{\mathbf{a}})=\delta_{\mathbf{b}}^{\mathbf{a}%
}$, $\mathbf{a,b}=0,1,2,3$. The reciprocal basis of $\{\theta^{\mathbf{b}}\}$
is denoted $\{\theta_{\mathbf{a}}\}$ and we have $\theta_{\mathbf{a}}%
\cdot\theta_{\mathbf{b}}=\eta_{\mathbf{ab}}$. Also,
\begin{equation}
\mathbf{\nabla}_{e_{\mathbf{a}}}\theta^{\mathbf{b}}=-\omega_{\mathbf{a}%
}^{\mathbf{bc}}\theta_{\mathbf{c}} \label{12n}%
\end{equation}
Defining
\begin{equation}
\mathbf{\omega}_{\mathbf{e}_{\mathbf{a}}}=\frac{1}{2}\omega_{\mathbf{a}%
}^{\mathbf{bc}}\theta_{\mathbf{b}}\wedge\theta_{\mathbf{c}}, \label{12nn}%
\end{equation}
we have that for any $A_{p}\in\sec\bigwedge^{p}T^{\ast}M,$ $p=0,1,2,3,4$
\begin{equation}
\mathbf{\nabla}_{\mathbf{e}_{\mathbf{a}}}A_{p}=\partial_{\mathbf{e}%
_{\mathbf{a}}}A_{p}+\frac{1}{2}[\mathbf{\omega}_{\mathbf{e}_{\mathbf{a}}%
},A_{p}], \label{12nnn}%
\end{equation}
where $\partial_{\mathbf{e}_{\mathbf{a}}}$ is the Pfaff derivative, i.e., if
$A_{p}=\frac{1}{p!}A_{\mathbf{i}_{1}...\mathbf{i}_{p}}\theta^{\mathbf{i}%
_{1}...\mathbf{i}_{p}}$,%
\begin{equation}
\partial_{\mathbf{e}_{\mathbf{a}}}A_{p}:=\frac{1}{p!}\mathbf{e}_{\mathbf{a}%
}(A_{\mathbf{i}_{1}...\mathbf{i}_{p}})\theta^{\mathbf{i}_{1}...\mathbf{i}_{p}%
}. \label{pfaaf}%
\end{equation}

Eq.(\ref{12nnn}) is an important formula which is also valid for a
nonhomogeneous $A\in\sec\mathcal{C}\!\ell(M,\mathtt{g})$. It is proved, e.g.,
in \cite{moro,rodoliv2006}.

\subsection{Dirac Operator Associated to a Levi-Civita Connection}

Using Eq.(\ref{12nnn}) we can show the very important result which is valid
for the Dirac operator associated to a \textit{Levi-Civita} connection denoted
${\mbox{\boldmath$\partial$}}$ :%

\begin{align}
{\mbox{\boldmath$\partial$}}A_{p}  &  ={\mbox{\boldmath$\partial$}}\wedge
A_{p\,}+\,{\mbox{\boldmath$\partial$}}\mathbin\lrcorner A_{p}=dA_{p}-\delta
A_{p},\nonumber\\
{\mbox{\boldmath$\partial$}}\wedge A_{p}  &  =dA_{p},\hspace{0.1in}%
\,{\mbox{\boldmath$\partial$}}\mathbin\lrcorner A_{p}=-\delta A_{p}.
\label{13}%
\end{align}

\section{Maxwell Theory in the Clifford Bundle}

With these results, Maxwell equations for $F\in\sec\bigwedge^{2}T^{\ast
}M\hookrightarrow\sec\mathcal{C}\!\ell(M,\mathtt{g})$, $J\in\sec\bigwedge
^{1}T^{\ast}M\hookrightarrow\sec\mathcal{C}\!\ell(M,\mathtt{g})$ reads%
\begin{equation}
dF=0,\text{ }\delta F=-J, \label{maxwell1}%
\end{equation}
or Maxwell equation\footnote{No misprint here.} reads (in a Lorentzian
spacetime)%
\begin{equation}
{\mbox{\boldmath$\partial$}}F=J{.} \label{maxwell2}%
\end{equation}

\subsection{Energy-Momentum Densities $\star\mathcal{T}_{\mathbf{a}}$ for the
Electromagnetic Field}

In this Appendix, we present a suggestive formula for the energy-momentum
densities $\star T_{\mathbf{a}}=-\star\mathcal{T}_{\mathbf{a}}$ of the Maxwell
field, namely:
\begin{equation}
\star\mathcal{T}_{\mathbf{a}}=-\frac{1}{2}\star(F\theta_{\mathbf{a}}\tilde
{F}). \label{emtem}%
\end{equation}
We also show that $\mathcal{T}_{\mathbf{a}}\cdot\theta_{\mathbf{b}%
}=\mathcal{T}_{\mathbf{b}}\cdot\theta_{\mathbf{a}}$. The derivation of those
formulas illustrates the power of the Clifford bundle formalism. In particular
\ref{emtem} simply cannot be written in the usual formalism of differential forms.

\hspace{-0.5cm}\emph{ }The Maxwell Lagrangian, here considered as the matter
field coupled to the background gravitational field must be taken (due to our
convention for the Ricci tensor and the definition of $\star\mathcal{T}%
_{\mathbf{a}}$ ) as%
\begin{equation}
\mathcal{L}_{m}=-\frac{1}{2}F\wedge\star F,
\end{equation}
where $F=\frac{1}{2}F_{\mathbf{ab}}\theta^{\mathbf{a}}\wedge\theta
^{\mathbf{b}}=\frac{1}{2}F_{\mathbf{ab}}\theta^{\mathbf{ab}}\in\sec%
{\displaystyle\bigwedge\nolimits^{2}}
TM\hookrightarrow\sec\mathcal{C\ell}(M,\mathtt{g})$ is the electromagnetic
field. \ We recall (as it is easy to verify) that
\[%
\mbox{\boldmath{$\delta$}}%
\star\theta^{\mathbf{ab}}=%
\mbox{\boldmath{$\delta$}}%
\theta^{\mathbf{c}}\wedge\lbrack\theta_{\mathbf{c}}\lrcorner\star
\theta^{\mathbf{ab}}].
\]
Also, for any \ $A_{p}\in\sec%
{\displaystyle\bigwedge\nolimits^{p}}
TM\hookrightarrow\sec\mathcal{C\ell}(M,\mathtt{g})$ we have
\begin{align*}
\lbrack%
\mbox{\boldmath{$\delta$}}%
,\star]A_{p}  &  =%
\mbox{\boldmath{$\delta$}}%
\star A_{p}-\star%
\mbox{\boldmath{$\delta$}}%
A_{p}\\
&  =%
\mbox{\boldmath{$\delta$}}%
\theta^{\mathbf{a}}\wedge(\theta_{\mathbf{a}}\lrcorner\star A_{p})-\text{
}\star\lbrack%
\mbox{\boldmath{$\delta$}}%
\theta^{\mathbf{a}}\wedge(\theta_{\mathbf{a}}\lrcorner A_{p})]
\end{align*}
Multiplying both members of the last equation with $A_{p}=F$ on the right by
$F\wedge$ we get%
\[
F\wedge%
\mbox{\boldmath{$\delta$}}%
\star F=F\wedge\star%
\mbox{\boldmath{$\delta$}}%
F+F\wedge\{%
\mbox{\boldmath{$\delta$}}%
\theta^{\mathbf{a}}\wedge(\theta_{\mathbf{a}}\lrcorner\star F)-\star\lbrack%
\mbox{\boldmath{$\delta$}}%
\theta^{\mathbf{a}}\wedge(\theta_{\mathbf{a}}\lrcorner F)]\}.
\]

Next we sum \ $%
\mbox{\boldmath{$\delta$}}%
F\wedge\star F$ to both members of the above equation obtaining%

\[%
\mbox{\boldmath{$\delta$}}%
\left(  F\wedge\star F\right)  =2%
\mbox{\boldmath{$\delta$}}%
F\wedge\star F+%
\mbox{\boldmath{$\delta$}}%
\theta^{\mathbf{a}}\wedge\lbrack F\wedge(\theta_{\mathbf{a}}\lrcorner\star
F)-(\theta_{\mathbf{a}}\lrcorner F)\wedge\star F].
\]
or,%
\[%
\mbox{\boldmath{$\delta$}}%
\left(  -\frac{1}{2}F\wedge\star F\right)  =-%
\mbox{\boldmath{$\delta$}}%
F\wedge\star F-\frac{1}{2}%
\mbox{\boldmath{$\delta$}}%
\theta^{\mathbf{a}}\wedge\lbrack F\wedge(\theta_{\mathbf{a}}\lrcorner\star
F)-(\theta_{\mathbf{a}}\lrcorner F)\wedge\star F].
\]
It follows that if $%
\mbox{\boldmath{$\delta$}}%
\theta^{\mathbf{a}}=-\pounds _{\xi}\theta^{\mathbf{a}}$ for some diffemorphism
generated by the vector field $\xi$, then
\[
\star\mathcal{T}_{\mathbf{a}}=\frac{\partial\mathcal{L}_{m}}{\partial
\theta^{\mathbf{a}}}=-\frac{1}{2}\left[  F\wedge(\theta_{\mathbf{a}}%
\lrcorner\star F)-(\theta_{\mathbf{a}}\lrcorner F)\wedge\star F\right]  .
\]
Now,
\[
(\theta_{\mathbf{a}}\lrcorner F)\wedge\star F=-\star\lbrack(\theta
_{\mathbf{a}}\lrcorner F)\lrcorner F]=-[(\theta_{\mathbf{a}}\lrcorner
F)\lrcorner F]\tau_{\mathtt{\mathbf{g}}}%
\]
and we have
\[
(\theta_{\mathbf{a}}\lrcorner F)\wedge\star F=\theta_{\mathbf{a}}(F\cdot
F)\tau_{\mathtt{\mathbf{g}}}-F\wedge(\theta_{\mathbf{a}}\lrcorner\star F).
\]
Using these results, we can write%
\begin{align*}
\frac{1}{2}\left[  F\wedge(\theta_{\mathbf{a}}\lrcorner\star F)-(\theta
_{\mathbf{a}}\lrcorner F)\wedge\star F\right]   &  =\frac{1}{2}\left\{
\theta_{\mathbf{a}}(F\cdot F)\tau_{\mathtt{\mathbf{g}}}-(\theta_{\mathbf{a}%
}\lrcorner F)\wedge\star F-(\theta_{\mathbf{a}}\lrcorner F)\wedge\star
F\right\} \\
&  =\frac{1}{2}\left\{  \theta_{\mathbf{a}}(F\cdot F)\tau_{\mathtt{\mathbf{g}%
}}-2(\theta_{\mathbf{a}}\lrcorner F)\wedge\star F\right\} \\
&  =\frac{1}{2}\left\{  \theta_{\mathbf{a}}(F\cdot F)\tau_{\mathtt{\mathbf{g}%
}}+2[(\theta_{\mathbf{a}}\lrcorner F)\lrcorner F]\tau_{\mathtt{\mathbf{g}}%
}\right\} \\
&  =\star\left(  \frac{1}{2}\theta_{\mathbf{a}}(F\cdot F)+(\theta_{\mathbf{a}%
}\lrcorner F)\lrcorner F\right)  =\frac{1}{2}\star(F\theta_{\mathbf{a}}%
\tilde{F}),
\end{align*}
where in writing the last line we used the identity
\begin{equation}
\frac{1}{2}Fn\tilde{F}=(n\lrcorner F)\lrcorner F+\frac{1}{2}n(F\cdot F),
\label{6.66a}%
\end{equation}
whose proof is as follows:
\begin{align*}
(n\lrcorner F)\lrcorner F+\frac{1}{2}n(F\cdot F)  &  =\frac{1}{2}\left[
(n\lrcorner F)F-F(n\lrcorner F)\right]  +\frac{1}{2}n(F\cdot F)\\
&  =\frac{1}{4}\left[  nFF-FnF-FnF+FFn\right]  +\frac{1}{2}n(F\cdot F)\\
&  =-\frac{1}{2}FnF+\frac{1}{4}\left[  -2n(F\cdot F)+n(F\wedge F)+(F\wedge
F)n\right]  +\frac{1}{2}n(F\cdot F)\\
&  =-\frac{1}{2}FnF+-\frac{1}{2}n(F\cdot F)+\frac{1}{2}n\wedge(F\wedge
F)+\frac{1}{2}n(F\cdot F)\\
&  =-\frac{1}{2}FnF=\frac{1}{2}Fn\tilde{F}.
\end{align*}
valid for any $n\in\sec%
{\displaystyle\bigwedge\nolimits^{1}}
T^{\ast}M\hookrightarrow\sec\mathcal{C\ell}(M,\mathtt{g})$ \ and \ $F$
$\in\sec%
{\displaystyle\bigwedge\nolimits^{2}}
T^{\ast}M\hookrightarrow\sec\mathcal{C\ell}(M,\mathtt{g})$.

(b) To prove that $\mathcal{T}_{\mathbf{a}}\cdot\theta_{\mathbf{b}%
}=\mathcal{T}_{\mathbf{b}}\cdot\theta_{\mathbf{a}}$ we write:
\begin{align*}
\mathcal{T}_{\mathbf{a}}\cdot\theta_{\mathbf{b}}  &  =\frac{1}{2}\langle
F\theta_{\mathbf{a}}F\theta_{\mathbf{b}}\rangle_{0}=\langle(F\llcorner
\theta_{\mathbf{a}})F\theta_{\mathbf{b}}\rangle_{0}+\frac{1}{2}\langle
(\theta_{\mathbf{a}}\lrcorner F\text{ }+\theta_{\mathbf{a}}\wedge F)\text{
}F\theta_{\mathbf{b}}\rangle_{0}\\
&  =\langle(F\llcorner\theta_{\mathbf{a}})F\theta_{\mathbf{b}}\rangle
_{0}+\frac{1}{2}\langle(\theta_{\mathbf{a}}FF\theta_{\mathbf{b}}\rangle_{0}\\
&  =\langle(F\llcorner\theta_{\mathbf{a}})(F\llcorner\theta_{\mathbf{b}%
})+(F\llcorner\theta_{\mathbf{a}})(F\wedge\theta_{\mathbf{b}})\rangle
_{0}-\frac{1}{2}\langle\theta_{\mathbf{a}}(F\cdot F)\theta^{\mathbf{b}}%
\rangle_{0}+\frac{1}{2}\langle\text{ }\theta_{\mathbf{a}}(F\wedge F)\text{
}\theta_{\mathbf{b}}\rangle_{0}\\
&  =\langle(F\llcorner\theta_{\mathbf{a}})(F\llcorner\theta_{\mathbf{b}%
})\rangle_{0}-\frac{1}{2}\langle(F\cdot F)(\theta_{\mathbf{a}}\cdot
\theta_{\mathbf{b}})\rangle_{0}\\
&  =(F\llcorner\theta_{\mathbf{b}})\cdot(F\llcorner\theta_{\mathbf{a}}%
)-\frac{1}{2}(F\cdot F)(\theta_{\mathbf{b}}\cdot\theta_{\mathbf{a}%
})=\mathcal{T}_{\mathbf{b}}\cdot\theta_{\mathbf{a}}.
\end{align*}

Note moreover that%
\begin{equation}
-T_{\mathbf{ab}}=\mathcal{T}_{\mathbf{ab}}=\mathcal{T}_{\mathbf{a}}\cdot
\theta_{\mathbf{b}}=\eta^{\mathbf{cl}}F_{\mathbf{ac}}F_{\mathbf{bl}}-\frac
{1}{4}F_{\mathbf{cd}}F^{\mathbf{cd}}\eta_{\mathbf{ab}},
\end{equation}
a well known result.

\section{Examples of Killing Vector Fields That Do Not Satisfy Eq.(\ref{7.62}%
)}

\subsection{Teleparallel Schwarzschild spacetime}

The metric of teleparallel Schwarzschild spacetime in spherical coordinates
is
\begin{equation}
\mathbf{g}=\zeta^{2}dt\otimes dt-\zeta^{-2}dr\otimes dr-r^{2}d\theta\otimes
d\theta-r^{2}\sin\theta d\phi\otimes d\phi,
\end{equation}
with%
\begin{equation}
\zeta:=\left(  1-\frac{k}{r}\right)  ^{1/2},
\end{equation}
\noindent where $k$ is a constant.

The Killing vector fields of this metric are

\begin{center}%
\begin{tabular}
[c]{||r||r|r|r|r||}\hline\hline
$p$ & $\xi^{0}$ & $\xi^{1}$ & $\xi^{2}$ & $\xi^{3}$\\\hline\hline
(1) & 1 & 0 & 0 & 0\\\hline
(2) & 0 & 0 & $-\sin\phi$ & $-\cot\theta\cos\phi$\\\hline
(3) & 0 & 0 & $\cos\phi$ & $-\cot\theta\sin\phi$\\\hline
(4) & 0 & 0 & 0 & 1\\\hline\hline
\end{tabular}
\medskip

\noindent{\footnotesize Table 1: Killing vectors associated with Schwarzschild
metric.}
\end{center}

\medbreak Introducing the orthonormal \ basis $\{\mathbf{e}_{\mathbf{a}}%
\}\in\sec P_{\mathrm{SO}_{1,3}^{e}}(M)$, where
\begin{equation}
\mathbf{e}_{0}=\zeta^{-1}\partial_{t},\quad\mathbf{e}_{1}=\zeta\partial
_{r},\quad\text{ }\mathbf{e}_{2}=\frac{1}{r}\partial_{\theta}\text{ }%
,\quad\text{ }\mathbf{e}_{3}=\frac{1}{r\sin\theta}\partial_{\phi},
\end{equation}
we get the for the structure coefficients of the basis (which are equal the
negative of the components of the torsion tensor in this basis),%
\[
c_{10}^{0}=-k\zeta^{-1}/r^{2},\quad c_{12}^{2}=\zeta/r=c_{13}^{3},\quad
c_{23}^{3}=\cot\theta/r.
\]

We then can verify that only the fourth Killing vector field in Table 1
satisfy Eq.(\ref{7.62}).

\subsection{Teleparallel de Sitter spacetime}

The metric of de Sitter teleparallel spacetime in spherical coordinate is for
$\alpha<\sqrt{R}$:%

\begin{equation}
\mathbf{g}=\omega^{2}dt\otimes dt-\omega^{2}dr\otimes dr-r^{2}\sin\theta
d\phi\otimes d\phi, \label{k1}%
\end{equation}
where
\begin{equation}
\omega:=(1-\alpha r^{2})^{\frac{1}{2}}, \quad\alpha=3/R^{2}, \label{k2}%
\end{equation}
with $\alpha$ the cosmological constant and $R$ the curvature radius. The ten
Killing vector fields of the de Sitter metric are ($c=\cosh(\sqrt{\alpha}t)$
and$s=\sinh(\sqrt{\alpha}t)$){\footnotesize ,}%

\begin{tabular}
[c]{||r||r|r|r|r||}\hline\hline
$p$ & $\xi^{0}$ & $\xi^{1}$ & $\xi^{2}$ & $\xi^{3}$\\\hline\hline
(1) & $r\omega^{-1}\sin\theta\cos\phi\;c$ & $\sqrt{\alpha}\omega\sin\theta
\cos\phi\;s$ & $\frac{\sqrt{\alpha}}{r}\omega\cos\theta\cos\phi\;s$ &
$-\frac{\sqrt{\alpha}}{r}\omega\frac{\sin\phi}{\sin\theta}\;s$\\\hline
(2) & $r\omega^{-1}\sin\theta\sin\phi\;c$ & $\sqrt{\alpha}\sin\theta\sin
\phi\;s$ & $\frac{\sqrt{\alpha}}{r}\omega\cos\theta\sin\phi\;s$ &
$-\frac{\sqrt{\alpha}}{r}\omega\frac{\cos\phi}{\sin\theta}\;s$\\\hline
(3) & $r\omega^{-1}\cos\theta\;c$ & $-\sqrt{\alpha}\omega\cos\theta\;s$ &
$-\frac{\sqrt{\alpha}}{r}\omega\sin\theta\;s$ & 0\\\hline
(4) & $-r\omega^{-1}\sin\theta\cos\phi\;s$ & $-\sqrt{\alpha}\omega\sin
\theta\cos\phi\;c$ & $-\frac{\sqrt{\alpha}}{r}\omega\cos\theta\cos\phi\;c$ &
$\frac{\sqrt{\alpha}}{r}\omega\frac{\sin\phi}{\sin\theta}\;c$\\\hline
(5) & $-r\omega^{-1}\sin\theta\sin\phi\;s$ & $-\sqrt{\alpha}\omega\sin
\theta\sin\phi\;c$ & $-\frac{\sqrt{\alpha}}{r}\omega\cos\theta\sin\phi\;c$ &
$-\frac{\sqrt{\alpha}}{r}\omega\frac{\cos\phi}{\sin\theta}\;c$\\\hline
(6) & $-r\omega^{-1}\cos\theta\;s$ & $-\sqrt{\alpha}\omega\cos\theta\;c$ &
$\frac{\sqrt{\alpha}}{r}\omega\sin\theta\;c$ & 0\\\hline
(7) & $\sqrt{\alpha}$ & 0 & 0 & 0\\\hline
(8) & 0 & 0 & $-\cos\phi$ & $\cot\theta\sin\phi$\\\hline
(9) & 0 & 0 & $-\sin\phi$ & $-\cot\theta\cos\phi$\\\hline
(10) & 0 & 0 & 0 & -1\\\hline\hline
\end{tabular}

{\footnotesize Table 2. Killing vectors associated with de Sitter teleparallel
spacetime for }$r<\sqrt{\alpha}${\footnotesize . } \bigskip

Introducing the orthonormal basis $\{\mathbf{e}_{\mathbf{a}}\}\in\sec
P\mathrm{SO}_{1,3}^{e}(M)$, where
\begin{equation}
\mathbf{e}_{0}=\omega^{-1}\partial_{t},\quad\mathbf{e}_{1}=\omega\partial
_{r},\quad\text{ }\mathbf{e}_{2}=\frac{1}{r}\partial_{\theta}\text{ }%
,\quad\text{ }\mathbf{e}_{3}=\frac{1}{r\sin\theta}\partial_{\phi}, \label{k3}%
\end{equation}
we get that the non null structure coefficients of the basis (which are the
negative of the components of the torsion tensor in this basis)
\begin{equation}
c_{10}^{0}=\alpha r\omega^{-1},\quad c_{12}^{2}=\omega/r=c_{13}^{3}%
,\quad\text{ }c_{23}^{3}=\cot\theta/r\text{.} \label{k4}%
\end{equation}
It can then be verified that only the seventh Killing vector field in Table 2
satisfy Eq.(\ref{7.62}).

When $r>\sqrt{\alpha}$ the metric of de Sitter teleparallel spacetime is%
\begin{equation}
\mathbf{g}=\Omega^{2}dt\otimes dt-\Omega^{2}dr\otimes dr-r^{2}\sin\theta
d\phi\otimes d\phi, \label{k5}%
\end{equation}
where
\begin{equation}
\Omega:=(\alpha r^{2}-1)^{\frac{1}{2}}\text{, }\alpha=3/R^{2},r>\sqrt{\alpha},
\label{k6}%
\end{equation}

As in the previous case, we have also ten Killing vector fields ($c=\cosh
(\sqrt{\alpha}t)$ and $s=\sinh(\sqrt{\alpha}t)$,%

\begin{tabular}
[c]{||r||r|r|r|r||}\hline\hline
$p$ & $\xi^{0}$ & $\xi^{1}$ & $\xi^{2}$ & $\xi^{3}$\\\hline\hline
1 & $r\Omega^{-1}\sin\theta\cos\phi\;s$ & $\sqrt{\alpha}\Omega\sin\theta
\cos\phi\;c$ & $\frac{\sqrt{\alpha}}{r}\Omega\cos\theta\cos\phi\;c$ &
$-\frac{\sqrt{\alpha}}{r}\Omega\frac{\sin\phi}{\sin\theta}\;c$\\\hline
2 & $r\Omega^{-1}\sin\theta\sin\phi\;s$ & $\sqrt{\alpha}\sin\theta\sin\phi\;c$
& $\frac{\sqrt{\alpha}}{r}\Omega\cos\theta\sin\phi\;c$ & $-\frac{\sqrt{\alpha
}}{r}\Omega\frac{\cos\phi}{\sin\theta}\;c$\\\hline
3 & $r\Omega^{-1}\cos\theta\;s$ & $-\sqrt{\alpha}\Omega\cos\theta\;c$ &
$-\frac{\sqrt{\alpha}}{r}\Omega\sin\theta\;c$ & 0\\\hline
4 & $-r\Omega^{-1}\sin\theta\cos\phi\;c$ & $-\sqrt{\alpha}\Omega\sin\theta
\cos\phi\;s$ & $-\frac{\sqrt{\alpha}}{r}\Omega\cos\theta\cos\phi\;s$ &
$\frac{\sqrt{\alpha}}{r}\Omega\frac{\sin\phi}{\sin\theta}\;s$\\\hline
5 & $-r\Omega^{-1}\sin\theta\sin\phi\;c$ & $-\sqrt{\alpha}\Omega\sin\theta
\sin\phi\;s$ & $-\frac{\sqrt{\alpha}}{r}\Omega\cos\theta\sin\phi\;s$ &
$-\frac{\sqrt{\alpha}}{r}\Omega\frac{\cos\phi}{\sin\theta}\;s$\\\hline
6 & $-r\Omega^{-1}\cos\theta\;c$ & $-\sqrt{\alpha}\Omega\cos\theta\;s$ &
$\frac{\sqrt{\alpha}}{r}\Omega\sin\theta\;s$ & 0\\\hline
7 & $\sqrt{\alpha}$ & 0 & 0 & 0\\\hline
8 & 0 & 0 & $-\cos\phi$ & $\cot\theta\sin\phi$\\\hline
9 & 0 & 0 & $-\sin\phi$ & $-\cot\theta\cos\phi$\\\hline
10 & 0 & 0 & 0 & -1\\\hline\hline
\end{tabular}
\medskip\noindent{\footnotesize Table 3: Killing vectors associated with de
Sitter teleparallel spacetime for }$r>\sqrt{\alpha}${\footnotesize . \bigskip}

Introducing the orthonormal basis $\{\mathbf{e}_{\mathbf{a}}\}\in\sec
P\mathrm{SO}_{1,3}^{e}(M)$, where
\begin{equation}
\mathbf{e}_{0}=\Omega^{-1}\partial_{t},\quad\mathbf{e}_{1}=\Omega\partial
_{r},\quad\text{ }\mathbf{e}_{2}=\frac{1}{r}\partial_{\theta}\text{ }%
,\quad\text{ }\mathbf{e}_{3}=\frac{1}{r\sin\theta}\partial_{\phi},
\end{equation}
we get once again the non null structure coefficients of the basis (which are
now the negative of the components of the torsion tensor in this basis)
\begin{equation}
c_{10}^{0}=\alpha r\Omega^{-1},\quad c_{12}^{2}=\Omega/r=c_{13}^{3}%
,\quad\text{ }c_{23}^{3}=\cot\theta/r\text{.}%
\end{equation}

It can then be verified that only the seventh Killing vector field in Table 3
satisfy Eq.(\ref{7.62}).

\subsection{Teleparallel Friedmann Spacetime}

Consider the metric of the following particular Friedmann spacetime in
comoving coordinates%
\[
\mathbf{g}=dt\otimes dt-R^{2}(t)(dx\otimes dx+dy\otimes dy+dz\otimes dz)
\]

\begin{tabular}
[c]{||r||r|r|r|r||}\hline\hline
$p$ & $\xi^{0}$ & $\xi^{1}$ & $\xi^{2}$ & $\xi^{3}$\\\hline\hline
(1) & 0 & 1 & 0 & 0\\\hline
(2) & 0 & 0 & 1 & 0\\\hline
(3) & 0 & 0 & 0 & 1\\\hline
(4) & 0 & $-y$ & $x$ & 0\\\hline
(5) & 0 & 0 & $z$ & $y$\\\hline
(6) & 0 & $z$ & 0 & $-x$\\\hline\hline\hline
\end{tabular}
\medskip

\noindent{\footnotesize Table 4: Killing vectors associated with Friedmann
metric. \bigskip} \medbreak

We see that there is no timelike Killing vector field. Introducing the
orthonormal basis $\{\mathbf{e}_{\mathbf{a}}\}\in\sec P_{\mathrm{SO}_{1,3}%
^{e}}(M)$, where
\begin{equation}
\mathbf{e}_{0}=\partial_{t},\quad\mathbf{e}_{1}=R^{-1}\partial_{x}%
,\quad\mathbf{e}_{2}=R^{-1}\partial_{y},\quad\mathbf{e}_{3}=R^{-1}\partial
_{z}.
\end{equation}
The non null structure coefficients of this basis (which are the negative of
the components of the torsion tensor in this basis) are
\begin{equation}
c_{10}^{0}=c_{20}^{2}=c_{30}^{3}=R^{-1}\dot{R}.
\end{equation}
and it can be verified that all Killing vector fields in Table~4 satisfy
Eq.(\ref{7.62}).

\section*{Acknowledgement}

Rold\~{a}o da Rocha thanks the Funda\c{c}\~{a}o de Amparo \`{a} Pesquisa do
Estado de S\~{a}o Paulo - Brazil (FAPESP) for financial support.


\begin{thebibliography}{99}                                                                                               %


\bibitem {pe1}{\footnotesize Aldrovandi R, Pereira J G, and Vu K H},
{\footnotesize Selected Topics in Teleparallel Gravity, \textit{Braz. J.
Phys.} \textbf{34}}, {\footnotesize 1374-1380 (2004) [\texttt{gr-qc/0312008]}}.

\bibitem {pe2}{\footnotesize Obukhov Yu N and Pereira J G},
{\footnotesize Metric-affine approach to teleparallel gravity, \textit{ Phys.
Rev. D} \textbf{67}}, {\footnotesize 044016 (2003) [\texttt{gr-qc/0212080]}}.

\bibitem {pe3}{\footnotesize de Andrade V C, Guillen L C T, and Pereira J G},
{\footnotesize Teleparallel Spin Connection, \textit{ Phys. Rev. D}
\textbf{64}}, {\footnotesize 027502 (2001) [\texttt{gr-qc/0104102]}}.

\bibitem {anderson}{\footnotesize Anderson, J. L., \textit{Principles of
Relativity Physics}, Academic Press, New York 1967.}

\bibitem {adm}{\footnotesize Arnowitt R. , Deser S., and Misner, C.},
{\footnotesize Coordinate Invariance and Energy Expressions in General
Relativity, \textit{Phys. Rev.} \textbf{122}}, {\footnotesize 997-1006 (1961)
[\texttt{gr-qc/0405109]}}.

\bibitem {benn}{\footnotesize Benn, I. M., Conservation Laws in Arbitrary
Space-times, \textit{Ann. Inst. H. Poincar\'{e}}, \textbf{XXXVII}, 67-91
(1982).}

\bibitem {benntucker}{\footnotesize Benn, I. M. and Tucker, R. W., \textit{An
Introduction to Spinors and Geometry}, Adam Hilger, Bristol and New York
1987.}

\bibitem {boro}{\footnotesize Bozhkov, Y., and Rodrigues, W. A. Jr., Mass and
Energy in General Relativity, \textit{ Gen. Rel. and Grav.}\textbf{ 27},
813-819 (1995).}

\bibitem {bramson}{\footnotesize Bramson, B. D., Relativistic Angular Momentum
for Asymptotically Flat Einstein-Maxwell Manifolds, \textit{Proc. R. Soc.
London Ser. A} \textbf{341}, 463-469 (1975).}

\bibitem {brown}{\footnotesize Brown, J. D. and York, J. W., Quasilocal Energy
and Conserved Charges Derived from the Gravitational Action, \textit{Phys.
Rev. D} \textbf{47}, 1407-1419 (1993).}

\bibitem {dalton}{\footnotesize Dalton, K., Energy and Momentum in General
Relativity, \textit{Gen. Rel. Grav}. \textbf{21}, 533-544 (1989)}

\bibitem {deandrade}{\footnotesize de Andrade, V. C., Guillen, L. C. T., and
Pereira, J. G., Gravitational Energy-Momentum Density in Teleparallel Gravity,
\textit{Phys. Rev. Lett. }\textbf{84}, 4533-4536 (2000).}

\bibitem {davis}{\footnotesize Davis, W. R., \textit{Classical Fields,
Particles and the Theory of Relativity}, Gordon and Breach, New York 1970.}

\bibitem {roldaowa}{\footnotesize da} {\footnotesize Rocha, R. and}
{\footnotesize Rodrigues, W.A. Jr., \ Diffeomorphism Invariance and Local
Lorentz Invariance, in Angl\`{e}s, P. and Jadczyk, A. (eds.), Proc. VII Int.
Conf. Clifford Algebras and their Applications, Toulouse 2005, Birkh\"{a}user,
Basel 2006 [\texttt{math-ph/0510026}]}.

\bibitem {einstein}{\footnotesize Einstein, A., Die Grundlage der Allgemeinen
Relativit\"{a}tstheorie, \textit{Ann. d. Phys}. \textbf{49}, 769-822 (1916).
[http://www.physik.uni-augsburg.de/annalen/history/papers/1916\_49\_769-822.pdf]
}

\bibitem {east}{\footnotesize Eastbrook, F. B., \textit{Conservation Laws for
Vacuum Tetrad Gravity} [\texttt{gr-qc/0508081}]}.

\bibitem {east2}{\footnotesize Eastbrook, F. B., Mathematical Structure of
Tetrad Equations for Vacuum Relativity, \textit{Phys. Rev. D }\textbf{71},
044004 (2005).}

\bibitem {Feynman}{\footnotesize Feynman, R. P., Morinigo, F. B. and Wagner,
W. G., (edited by Hatfield, B.), \textit{Feynman Lectures on Gravitation},
Addison-Wesley Publ. Co., Reading, MA 1995.}

\bibitem {franferr}{\footnotesize Francaviglia, M. and Ferraris, M.,
Energy-Momentum Tensors in Geometric Field Theories, \textit{J. Math. Phys.}
\textbf{26}, 1243-1252 (1965).}

\bibitem {freud}{\footnotesize Freud, P., \"{U}ber die Ausdr\"{u}cke der
Gesamtenergie und des Gesamtimpulses eines Materiellen Systems in der
Allgemeinen Relativit\"{a}tstheorie, \textit{Ann. Math}. \textbf{40}, 417-419
(1939).}

\bibitem {geroch}{\footnotesize Geroch, R. Spinor Structure of Space-Times in
General Relativity I, \textit{J. Math. Phys.} \textbf{9}, 1739-1744 (1968).}

\bibitem {gronwaldhehl}{\footnotesize Gronwald, F. and Hehl, F. W., On the
Gauge Aspects of Gravity, in Bergmann, P. G., P. G., de Sabatta, V. and
Treder, H.-J. (eds.), \ \textit{ Int. School of Cosmology and Gravitation:
14}}$^{th}\mathit{\ }$ {\footnotesize \textit{Course: Quantum Gravity,}May
1995, Erice, Italy, The Science and Culture Series 10, 148-198, World Sci.
Publ., River Edge, NJ 1996 [\texttt{gr-qc/9602013}]}.

\bibitem {hawellis}{\footnotesize Hawking, S. W. and Ellis, G. F. R.,
\textit{The Large Scale Structure of Spacetime}, Cambridge University Press,
Cambridge 1973.}

\bibitem {haya}{\footnotesize Hayashi, K. and Shirafuji, T., New General
Relativity, \textit{Phys. Rev. D} \textbf{19}, 3542-3553 (1979).}

\bibitem {hehl}{\footnotesize Hehl, F. W., \ von der Heyde, P., and Kerlick,
G. D., General Relativity with Spin and Torsion: Foundations and Prospects,
\textit{Rev. Mod. Phys}. \textbf{48}, 393-416 (1976).}

\bibitem {kop}{\footnotesize Kopczynski, W., Problems with metric-teleparallel
theories of gravitation, \textit{J. Phys. A}\textbf{15}, 493 (1982).}

\bibitem {lawmi}{\footnotesize Lawson, H. Blaine, Jr. and Michelson, M. L.,
\textit{Spin Geometry}, Princeton University Press, Princeton 1989.}

\bibitem {logunov1}{\footnotesize Logunov, A. A., Mestvirishvili, \textit{The
Relativistic Theory of Gravitation}, Mir Publ., Moscow 1989.}

\bibitem {logunov2}{\footnotesize Logunov, A. A,\textit{ Relativistic Theory
of }G\textit{ravity}, Nova Science Publ., New York 1999.}

\bibitem {maluf}{\footnotesize Maluf, J. W., Hamiltonian Formulation of the
Teleparallel Description of General Relativity, \textit{J. Math. Phys.
}\textbf{35}\textit{, }335-343 (1994).}

\bibitem {maluf2}{\footnotesize Maluf, J. W., Localization of Energy in
General Relativity, \textit{J. Math. Phys.} \textbf{36}, 4242-4247 (1995)}.

\bibitem {miel}{\footnotesize Mielke E. W,E. W. Mielke, "Generating Function
for Ashtekar's Complex Variables in General Relativity," \emph{Ann. Phys.}
\textbf{219}, 78-108 (1992).}

\bibitem {mingu}{\footnotesize Minguzzi, E., Gauge Invariance in Telparallel
Gravity Theories: A Solution to the background Structure Problem,
\textit{Phys. Rev. D }\textbf{65}, 084048 (2002).}

\bibitem {mtw}{\footnotesize Misner, C. M., Thorne, K. S. and Wheeler, J. A.,
\textit{Gravitation}, W.H. Freeman and Co. San Francisco 1973.}

\bibitem {moller}{\footnotesize M\o ller, C., On the Localization of the
Energy of a Physical System in the General Theory of Relativity \textit{Ann.
Phys.} \textbf{4}, 347-461 (1958).}

\bibitem {moller2}{\footnotesize M\o ller, C., Further Remarks on the
Localization of the Energy in the General Theory of Relativity \textit{Ann.
Phys.} \textbf{12}, 118-133 (1958).}

\bibitem {moller3}{\footnotesize M\o ller, C., Conservation Laws and Absolute
Parallelism in General Relativity, \textit{Mat.-Fys. Skr. K. Danske Vid.
Selsk} \textbf{1}, 1-50 (1961).}

\bibitem {moro}{\footnotesize Mosna, R. A. and Rodrigues, W. A., Jr., The
Bundles of Algebraic and Dirac-Hestenes Spinor Fields, \textit{J. Math. Phys}
\textbf{45}, 2945-2966 (2004) [\texttt{math-ph/0212033}]}.

\bibitem {mol}{\footnotesize M\"{u}ller-Hoissen, F. and Nitsch, J., \emph{On
the tetrad theory of gravity}, \textit{Gen. Rel. Grav.}\textbf{17}, 747-760
(1985).}

\bibitem {murchada}{\footnotesize Murchada, N. O.,Total Energy Momentum in
General Relativity, \textit{J. Math. Phys}. \textbf{27}, 2111-2118 (1986).}

\bibitem {nakahara}{\footnotesize Nakahara M, \textit{Geometry, Topology and
Physics}, Institute of Physics Publ., Bristol and Philadelphia}
{\footnotesize 1990}.

\bibitem {nester}{\footnotesize Nester, J. M., Positive Energy Via the
Teleparallel Hamiltonian, \textit{Int. J. Mod. Phys. A} \textbf{4}, 1755-1772
(1989).}

\bibitem {rodrigues2006}{\footnotesize Notte-}{\footnotesize Cuello, E. A. and
Rodrigues, W. A. Jr., A Maxwell Like Formulation of Gravitational Theory in
Minkowski Spacetime, accepted for publ. \ in \textit{Int. J. Mod. Phys. D
}(2006), [\texttt{math-ph/0608017}]}.

\bibitem {komar1}{\footnotesize Komar, A., Asymptotic Covariant Laws for
Gravitational Radiation, Phys. Rev. \textbf{127}, 1411-1418 (1962).}

\bibitem {komar2}{\footnotesize Komar, A., Positive-Definite Energy Density
and Global Consequences for General Relativity, \textit{Phys. Rev.}
\textbf{129}, 1873-1876 (1963).}

\bibitem {komar3}{\footnotesize Gravitational Superenergy as a Generator of
Canonical Transformation, \textit{Phys. Rev.} \textbf{164}, 1595-1599 (1967).}

\bibitem {komar4}{\footnotesize Komar, A., Generators of Coordinate
Transformations in the Penrosoe Formalism of General Relativity, \textit{Phys.
Rev}. \textbf{127}, 955-959 (1962).}

\bibitem {komar5}{\footnotesize Komar, A., Enlarged Gauge Symmetry of
Gravitational Radiation, \textit{Phys. Rev. }\textbf{30}, 305-308 (1984).}

\bibitem {penrose05}{\footnotesize Penrose, R., \textit{The Road to Reality: A
Complete Guide to the Laws of the Universe}, Knopf Publ., New York 2005.}

\bibitem {rod04}{\footnotesize Rodrigues, W. A. Jr., Algebraic and
Dirac-Hestenes Spinor and Spinor Fields. \textit{J. Math. Phys.} \textbf{45},
2908-2945 (2004).[math-ph/0212030]}

\bibitem {rodroldao}{\footnotesize Rodrigues, W.A. Jr., da Rocha R., and Vaz,
J. Jr., Hidden Consequence of Local Lorentz Invariance, \textit{Int. J. Geom.
Meth. Mod. Phys.} \textbf{2}, 305-357 (2005) [\texttt{math-ph/0501064}]}.

\bibitem {roqui}{\footnotesize Rodrigues, W. A. Jr., and Souza, Q. A. G., The
Clifford Bundle and the Nature of the Gravitational Field, \textit{Found. of
Phys.} \textbf{23}, 1465--1490 (1993).}

\bibitem {roqui2005}{\footnotesize Rodrigues, W. A. Jr., and Souza, Q. A. G.,
An Ambigous Statement Called \ `Tetrad Postulate' and the Correct Field
Equations Satisfied by the Tetrad Fields, \textit{Int. J. Mod. Phys. D.
}\textbf{12}, 2095-2150 (2005). }

\bibitem {rodoliv04}{\footnotesize Rodrigues, W. A. Jr. and Oliveira, E.
Capelas,Clifford Valued Differential Forms, and Some Issues in Gravitation,
Electromagnetism and \ `Unified Theories', \textit{Int. J. Mod. Phys. D}
\textbf{13}, 1879-1915 (2004) [\texttt{math-ph/0407024}]. }

\bibitem {rodoliv2006}{\footnotesize Rodrigues, W. A. Jr. and Oliveira, E.
Capelas, \textit{The Many Faces of Maxwell, Dirac and Einstein Equations. A
Clifford Bundle Approach}, in publ. in Lecture Notes in Physics, Springer, New
York, 2007.}

\bibitem {sparling}{\footnotesize Sparling, G. A. J., \textit{Twistors,
Spinors and the Einstein Vacuum Equations} (unknown status), University of
Pittsburg preprint (1982).}

\bibitem {sawu}{\footnotesize Sachs, R. K., and Wu, H., General Relativity for
Mathematicians, Springer-Verlag, New York 1977. }

\bibitem {schoenyau1}{\footnotesize Schoen, R., and Yau, S.-T., Proof of the
Positive Mass Conjecture in General Relativity, \textit{Commun. Math. Phys.}
\textbf{65}, 45-76 (1979). }

\bibitem {schoenyau2}{\footnotesize Schoen, R., and Yau, S.-T., Proof of the
Positive Mass Theorem 2, \textit{Commun. Math. Phys.} \textbf{79}, 231-260
(1981). }

\bibitem {schwinger}{\footnotesize Schwinger, J., \textit{Particles, Sources
and Fields}, vol. 1, Addison-Wesley Publ. Co., Reading, MA 1970. }

\bibitem {szabados}{\footnotesize Szabados, L. B., Quasi-Local Energy-Momentum
and Angular Momentum in GR: A Review Article, \emph{Living Reviews in
Relativity}, } {\footnotesize [http://www.livingreviews.org/lrr-2004-4]}

\bibitem {thirring0}{\footnotesize Thirring, W., An Alternative Approach to
the Theory of Gravitation, \textit{Ann. Phys.} \textbf{16}, 96-117 (1961). }

\bibitem {thirring}{\footnotesize Thirring, W., \textit{A Course in
Mathematical Physic}s, vol.2, Springer-Verlag, New York 1979. }

\bibitem {thiwal}{\footnotesize Thirring, W. and Wallner, R., The Use of
Exterior Forms in Einstein's Gravitational Theory, \textit{Brazilian J. Phys.}
\textbf{8}, 686-723 (1978). }

\bibitem {trautman0}{\footnotesize Trautman, A., Conservation Laws in General
Relativity, in Witten, L.(ed.), \textit{Gravitation: An Introduction to
Current Research}, 169-198, J. Wiley \& Sons, Inc., New York, 1962. }

\bibitem {trautman1}{\footnotesize Trautman, A., On the Einstein--Cartan
Equations Part I, Bull. \textit{Acad. Polon. Sci., (S\'{e}r. Sci. Math., Astr.
et Phys.)} \textbf{20}, 185-190 (1972). }

\bibitem {trautman2}{\footnotesize Trautman, A., On the Einstein--Cartan
Equations, Part II, \textit{Bull. Acad. Polon. Sci., (S\'{e}r. Sci. Math.,
Astr. et Phys.)} \textbf{20}, 503-506 (1972). }

\bibitem {trautman3}{\footnotesize Trautman, A., On the Einstein--Cartan
Equations, Part III, \textit{Bull. Acad. Polon. Sci., (S\'{e}r. Sci. Math.,
Astr. et Phys.)} \textbf{20}, 895--896 (1972). }

\bibitem {trautman4}{\footnotesize Trautman, A., On the Einstein--Cartan
Equations, Part IV, \textit{Bull. Acad. Polon. Sci., (S\'{e}r. Sci. Math.,
Astr. et Phys.)} \textbf{21}, 345--346 (1973). }

\bibitem {trautman5}{\footnotesize Trautman, A., A Methaphysical Remark on
Variational Principles, \textit{Acta Phys. Polon. B} \textbf{27}, 839-847
(1996). }

\bibitem {trautman6}{\footnotesize Trautman, A., The Einstein-Cartan Theory,
in Francoise, J. P. , Naber, G. L. and Tsou, S. T. (eds.),
\textit{Encyclopedia of \ Mathematical Physics,} vol. 2., 189-195, Elsevier,
Amsterdam 2006 [\texttt{gr-qc/0606062 }]. }

\bibitem {tung}{\footnotesize Tung R. S. and Nester J. M., The quadratic
spinor Lagrangian is equivalent to teleparallel theory, \textit{Phys. Rev.
D}\textbf{60} 021501 (1999).}

\bibitem {vatorr1}{\footnotesize Vargas, J. G., and Torr, D. G., Conservation
of Vector-Valued Forms and the Question of the Existence of Gravitational
Energy-Momentum in General Relativity,\textit{ Gen. Rel.} \textit{Grav.}%
\textbf{ 23}, 713-732 (1991). }

\bibitem {wald}{\footnotesize Wald, R., \textit{General Relativity}, Univ.
Chicago Press, Chicago 1984. }

\bibitem {wallner}{\footnotesize Wallner, R. P., Notes on the Gauge Theory of
Gravitation, \textit{Acta Phys. Austriaca }\textbf{54}, 165-189 (1882). }

\bibitem {weinberg1}{\footnotesize Weinberg, S., Photons and Gravitons in
Pertubation Theory: Derivation of Maxwell's and Einstein's Equations,
\textit{Phys. Rev. B} \textbf{138}, 988-1002 (1965). }

\bibitem {witten}{\footnotesize Witten, E., \ A New Proof of the Positive
Energy Theorem, \textit{Comm. Math. Phys}. \textbf{80}, 381-402 (1981).}
\end{thebibliography}
\end{document}